\documentclass[11pt]{article}

\usepackage{amsfonts}
\usepackage{amsmath}
\usepackage{amssymb}
\usepackage{graphicx}
\usepackage{dcolumn}
\usepackage{dsfont,soul}
\usepackage{times}
\usepackage{epsfig,cite}
\usepackage[]{units}
\usepackage{todonotes}

\usepackage{lipsum} 

\usepackage[margin=1in]{geometry}

\newcommand{\R}{{\mathbb R}}

\newcommand{\Z}{{\mathbb Z}}

\newcommand{\C}{{\mathbb C}}

\newcommand{\eps}{\varepsilon}

\newcommand{\pa}{\partial}

\def\epsilon{\varepsilon}

\def\beq{\begin{equation}}
\def\eeq{\end{equation}}

\newcommand{\subfigimg}[3][,]{%
  \setbox1=\hbox{\includegraphics[#1]{#3}}
  \leavevmode\rlap{\usebox1}
  \rlap{\hspace*{5pt}\raisebox{\dimexpr\ht1-.5\baselineskip}{#2}}
  \phantom{\usebox1}
}

\newcommand{\jtitle}[1]{ \emph{#1} \kern-.5ex}

\begin{document} 
\title{Nonlinear Coherent Structures in Granular Crystals}
\author{C. Chong\footnote{Email: cchong@bowdoin.edu}, Mason A. Porter, P. G. Kevrekidis, C. Daraio }


\date{\today}


\maketitle

\begin{abstract}

The study of granular crystals, metamaterials that consist of closely packed arrays of particles that interact elastically, is a vibrant area of research that combines ideas from disciplines such as materials science, nonlinear dynamics, and condensed-matter physics. Granular crystals, a type of nonlinear metamaterial, exploit geometrical nonlinearities in their constitutive microstructure to produce properties (such as tunability and energy localization) that are not conventional to engineering materials and linear devices. In this topical review, we focus on recent experimental, computational, and theoretical results on nonlinear coherent structures in granular crystals. Such structures --- which include traveling solitary waves, dispersive shock waves, and discrete breathers --- have fascinating dynamics, including a diversity of both transient features and robust, long-lived patterns that emerge from broad classes of initial data. In our review, we primarily discuss phenomena in one-dimensional crystals, as most research has focused on such scenarios, but we also present some extensions to two-dimensional settings. Throughout the review, we highlight open problems and discuss a variety of potential engineering applications that arise from the rich dynamic response of granular crystals.

\end{abstract}

\tableofcontents



\section{Introduction}  \label{sec:intro}

Granular crystals consist of closely packed arrays of particles that are modeled traditionally as interacting with each other elastically\footnote{See Section \ref{subsec:model} for a brief discussion of more general types of interactions.}\cite{Nester2001,sen08,theocharis_review,ptpaper} (see Fig.~\ref{fig:exp}). The simplest description of a granular crystal, which one can construe as a type of metamaterial, models only forces that result from the contact of adjacent particles. The exact nature of this contact force depends on the geometry of the particles in contact, the contact angle, and 
elastic properties of the particles \cite{Johnson}. Consequently, by changing the material properties or shapes of the particles, one can modify the effective stiffness of a granular crystal. One can apply an initial prestress (i.e., a precompression) by applying a static force at the boundaries of the crystal. This controls the strength of the system's nonlinearity, which can range from almost linear (for strong precompression) to purely nonlinear (no precompression). The remarkable tunability of a granular crystal stems both from this ability to tune the response from a weakly nonlinear regime to a highly nonlinear regime and from the ability to study
crystals with either homogeneous or controllably heterogeneous configurations (e.g., periodic ones, settings with defects, random ones, and so on). This flexibility makes granular crystals potentially useful for a host of applications, including shock- and energy-absorbing layers~\cite{dar06,hong05,fernando,doney06}, actuating devices \cite{dev08}, acoustic lenses \cite{Spadoni}, acoustic diodes \cite{Nature11} and switches \cite{Li_switch}, and sound scramblers \cite{dar05,dar05b}.

Granular crystals are also an ideal test bed for fundamental studies of nonlinear dynamics, as they constitute a prototypical (and experimentally relevant) realization of a classical system of nonlinearly coupled oscillators. The simplest and most standard type of granular crystal consists of a one-dimensional (1D) crystal with spherical particles. One can envision these spheres as particles that are coupled by nonlinear springs and do not have any on-site forces. We use the term ``granular chain'' to refer to an array of particles aligned in a 1D configuration. More generally, the term ``granular crystals'' includes such chains as examples but can also indicate more general configurations. The broad range of interest in granular chains (and in granular crystals more generally) is apparent in the diverse nature of the publications about them; these range from predominantly experimental investigations that focus on applications to theoretical studies that concentrate on rigorous mathematical proofs. V.~F. Nesterenko is largely responsible for bringing significant attention to the subject of granular crystals, and his influential 2001 book \cite{Nester2001} is a canonical starting point for researchers who are interested in this subject. Since the publication of Nesterenko's book, there has been a burst of activity on granular crystals. These activities were discussed in a 2008 review article by S. Sen et al. \cite{sen08}, and specific areas in the study of granular crystals have been surveyed in specialized reviews that were published as book chapters \cite{theocharis_review,vakakis_review}. There is also a very recent popular article in the magazine \emph{Physics Today}~\cite{ptpaper}. Our review complements \cite{sen08,theocharis_review,vakakis_review} and highlights a wealth of exciting research on granular crystals has appeared since those papers were published.\footnote{We also note that although other types of granular systems --- such as soils, sand, and cereals --- also have rich and diverse dynamics, their physical properties are rather different from those of granular crystals. They are beyond the scope of our article. Interested readers can examine 
the reviews~\cite{jamming2015,jamming2010}.}

The explosion of recent results in granular crystals has resulted from the emergence of numerous research groups with the ability to perform detailed quantitative monitoring of granular crystals by using novel techniques (including noninvasive ones, such as photography \cite{l8} or laser Doppler vibrometry \cite{plastic2,Yang_laser2013}). These exciting advances have ushered
the investigation of  granular crystals into a new era of theory, computation, and experiments ~\cite{ptpaper} that we will explore in the present paper.

In this topical review, we emphasize nonlinear coherent structures, such as traveling solitary waves, dispersive shock waves,
and discrete nonlinear breathers. Although our article undoubtedly reflects our personal biases and research interests within this research theme, we hope to draw attention to a broad range of new results. 

The remainder of our topical review is structured as follows. We present experimental setups in Sec.~\ref{subsec:experiments}, discuss model equations in Sec.~\ref{subsec:model}, and examine the linear regime and dispersion in Sec.~\ref{sec:linear}. We then discuss traveling solitary waves in Sec.~\ref{sec:TW}, dispersive shock waves in Sec.~\ref{sec:shocks}, and breathers in Sec.~\ref{sec:breathers}. We give an introduction to two-dimensional granular crystals in Sec.~\ref{sec:2D}, and we conclude in Sec.~\ref{sec:theend}. We highlight open questions and possible applications of granular crystals throughout the article.


\section{Experimental Setups} \label{subsec:experiments}

 \begin{figure} 
\centerline{
 \includegraphics[width=.9 \linewidth]{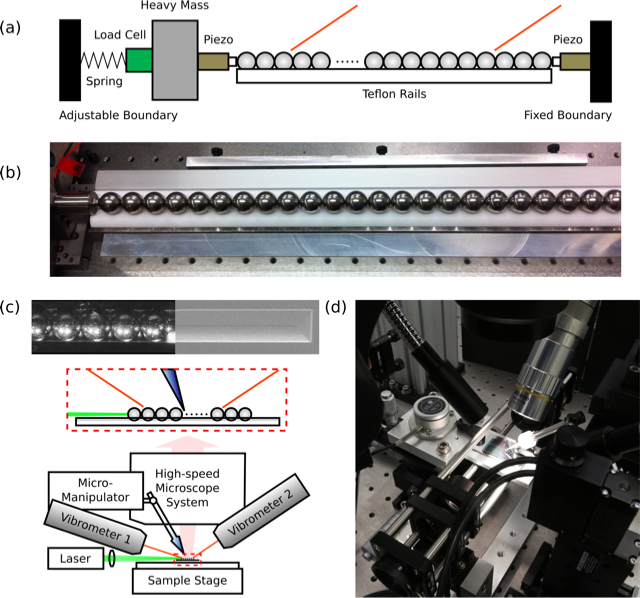}
  }
\caption{Macro- and microscopic experimental setups for testing nonlinear granular chains. \textbf{(a)} Schematic diagram of a typical macroscale testing system, including support rods, boundaries, excitation and detection system (the red lines here represent laser interferometers, monitoring the displacement and velocity of selected particles). \textbf{(b)} Digital image of (a). Here, the particles have a diameter of ca. 1.9 cm. \textbf{(c)} Schematic diagram of a typical microscale testing system, including support grooves, robotic manipulation, optical excitation and detection systems. The inset shows an optical image of the granular chains composed of stainless steel particles of ca. 300 microns diameter and a scanning electron microscopy image of the support grooves. \textbf{(d)} Digital image of the setup in (c). [We use panels (a,b) with permission from Paul Anzel. Panels (c,d) have been modified with permission from Ref.~\cite{Lin} and is copyrighted (2016) by the American Physical Society.]
 }
\label{fig:exp}
\end{figure}


Most experimental studies on the dynamic response of granular crystals have focused on the elastic response of one-dimensional (1D) and two-dimensional (2D) macroscopic systems, usually composed of 10--800 particles, with diameters in the millimeter-to-centimeter range. Most granular crystals use spherical particles, although an increasing number of recent studies have examined other geometries, including elliptical particles \cite{Ngo:PRE2011}, cylindrical particles \cite{khatri2012,kim-yang}, and a combination of different particle shapes \cite{l10}. Researchers have used particles made of various materials~\cite{Nester2001,sen08,theocharis_review}.
The most commonly used constitutive materials are stainless steel ball bearings~\cite{Nester2001}, although other metals (e.g., aluminum, brass, and bronze) \cite{Coste,Coste1999,Boechler2010} and softer crystals --- composed of polymeric particles such as Homalite \cite{shukla1993}, nylon \cite{Coste1999}, Teflon (PTFE) \cite{dar05}, and Delrin \cite{Andrea} --- have also been employed.

Experimental techniques employed for monitoring wave propagation through granular crystals (see Fig.~\ref{fig:exp}[a,b]) often assemble particles inside cylindrical \cite{dar05b} or square support guides \cite{Coste}, support rods \cite{Boechler2010}, or tracks \cite{Melo05}.
In these systems, the initial excitation of propagating waves has been created using drop-weight impacts of striker particles  \cite{dar05b}, electromagnetic actuators \cite{Coste}, piezo-electric transducers \cite{deBilly01}, and noncontact laser ablation \cite{Xianglei}. Typical initial excitations of granular chains include single impulses, shocks, and continuous harmonic vibrations. Much of the same experimental approaches used in 1D granular chains have also been employed to test nonlinear wave propagation in elastic 2D \cite{Leonard11,Andrea} and three-dimensional (3D) granular crystals with Hertzian contacts \cite{Gibiat}.

There are a variety of ways to detect waves propagating through granular crystals. The most common approach of a local measurement of transient force profiles in selected particles relies either on using piezo-electric transducers embedded at a boundary and/or inside the particles \cite{dar05,wetdry} or on using triaxial accelerometers \cite{Leonard11}. The use of embedded sensors allows the accurate detection of force profiles, but it is limited to a small number of locations in a granular crystal. Additionally, it can be intrusive to the dynamics from (i) an impedance mismatch between the piezoelectric layer and the particle materials and (ii) the presence of wires, which are needed to transfer the recorded signal. Less intrusive techniques include using laser vibrometers \cite{Yang_laser2013,plastic2}, which can measure the displacement and velocity of selected particles.
This approach has the advantage of the ability to scope individual beads at will, leading in principle to the potential to visualize the entire spatio-temporal dynamics of a granular crystal. Optical techniques, such as photo elasticity, have also enabled full-field visualization of propagating waves in 1D and 2D systems, although such approaches are limited to crystals composed of relatively soft particles and to disk-shaped elements (cylinders that are relatively small along the principle axis) \cite{l8,Glam}. Additionally, carbon paper has been employed to measure transmitted force at the edge of 3D particle lattices \cite{Mueggenburg}.

Experimental methods to test 1D granular crystals 
are now well-established. One of the challenges facing experimentalists today is the ability to create granular crystals in higher dimensions (in 2D and even 3D), where it is necessary to use a larger number of particles than in 1D. Increasing the number of particles in a highly packed crystal translates to increasing the amount of connectivity between particles. This demands high precision in particle fabrication (in particular, in their sphericity and in their surface finish and asperities), and small tolerances in crystal assembly (which, for the most part, have been arranged manually to date), both of which are difficult to achieve. Experimental systems with a very large number of particles are also more sensitive to the presence of defects (e.g., gaps between contacts or misalignment) and disorder, and such configurations are therefore less reproducible than systems with fewer particles \cite{l11}. 

An exciting experimental direction in granular crystals explored the dynamics of systems that are composed of ``engineered'' particles. For example, heavy particles with a soft coating \cite{daraio2006pre} have an exceptionally low signal speed, and particles with internal \cite{Daraio2015PRE} or external resonant masses \cite{gantzounis} allow the formation of tunable and low-frequency band gaps in transmission spectra. 

The experimental platforms that have been developed for granular crystals have yielded fundamental proof-of-principle demonstrations of the nonlinear phenomena predicted by computational and analytical studies. However, most of the experimental data is limited to the study of crystals with contact interactions in the linear elastic regime. Deviations from the purely elastic regime have been identified in the presence of dissipation \cite{rosas2007,khatriprl,brogliato2014} arising in experiments either from (1) the presence of friction between particles and supports or from (2) intrinsic losses in the particles (e.g., vibrational modes and viscoelasticity). Recent experimental approaches that deviate from linear elasticity have been helpful for characterizing granular crystals with elasto-plastic contacts. The effects of plasticity around a contact between particles is an important factor for the application of granular crystals as impact-protecting layers. Experimental setups to study impulse mitigation and redirection in elasto-plastic granular crystals excited by high-velocity impactors have included modified split-Hopkinson bar systems \cite{Burgoyne,Lambros1,Lambros2} and spring-loaded coils \cite{burgoyne2016}.

The studies mentioned above concern macroscopic particles (with millimeter-sized to centimeter-sized diameters). The solitary waves supported by nonlinear granular chains have a spatial wavelength of roughly five particle diameters (although they are not genuinely compact, as we will discuss in Secs.~\ref{sec:qc} and \ref{sec:genuineTW}) and a propagation speed that is proportional to their amplitude. These spatial and temporal properties translate to signals that travel through macroscopic granular chains with frequencies in the 1--100 kHz range. Using highly nonlinear signals in acoustic imaging or in nondestructive evaluation requires acoustic excitations in the ultrasonic regime (i.e., microscopic particle sizes). Although recent numerical investigations have examined wave propagation through granular crystals of smaller scales (e.g., nanoscale buckyball particle chains \cite{Buckyball}), much remains unknown about the propagation of dynamic excitations through chains of micrometer-size particles in contact. Although one expects the fundamental elasticity of the contacts is to remain valid when particles have diameters of about 1--100 micrometers, the dynamics of microparticle chains are not simply scaled-down version of their macroscale counterparts. One expects microscopic granular crystals 
to exhibit qualitatively different dynamics, which are affected by the presence of adhesive forces, electrostatic charges, surface asperities, and a greater sensitivity to disorder. Because of these effects, experimental studies at these scales are challenging: they require high accuracy in particle packing, excitation, and detection of waves with noncontact techniques (see the next paragraph), which are nonintrusive and more precise than those that are used for macroscale systems.

Recent studies of 1-micrometer diameter silica spheres employed a laser-induced transient grating technique to excite surface acoustic waves and measure their dispersion \cite{Fang}. However, this technique is not suitable for capturing nonlinear, transient wave phenomena, such as those found in macroscopic systems. At slightly larger scales, chains of steel particles with diameters of  about 2--300 micrometers have been investigated using noncontact optical techniques \cite{Lin}. In these experiments, particles were aligned in microfabricated grooves using a robotic positioning systems and relied on high-speed microscopy imaging for high-accuracy in positioning (see Fig.~\ref{fig:exp}[c,d]). 

\section{Model Equations} \label{subsec:model}

\begin{figure} 
    \centering
      \includegraphics[width=.6 \linewidth]{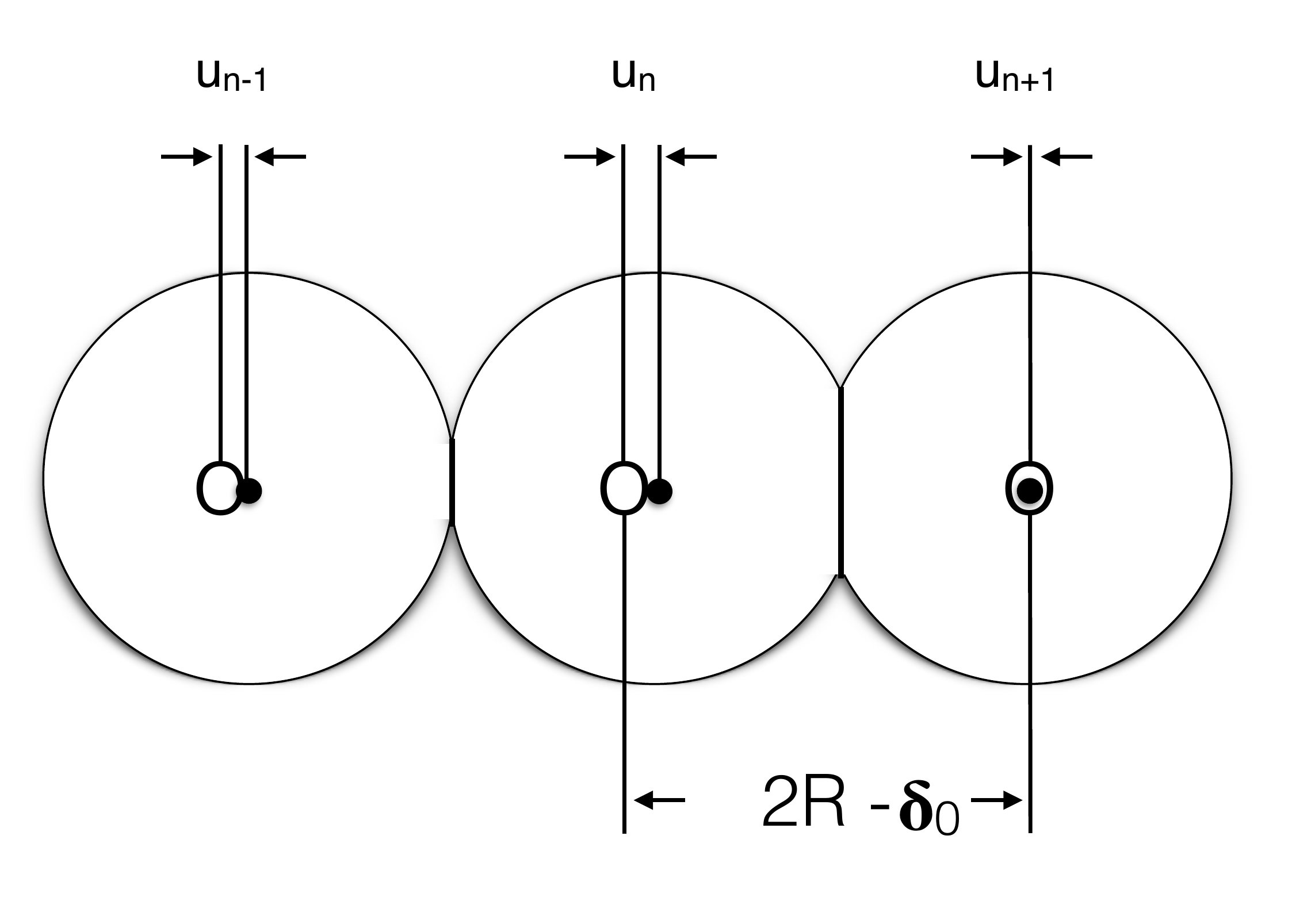} 
\caption{
One-dimensional monomer granular chain compressed by a static force $F_0$. The open circles represent the initial positions of the particle centers in static equilibrium, and the black circles indicated the displaced positions.  
[This figure was inspired by~Ref.~\cite{Nester2001}.]
}
\label{fig:Hertz}
\end{figure}

The force resulting from the deformation of two spherical particles in contact is
described by the classical Hertz law 
\begin{equation}
	F(x) = A [x]_+^{3/2}\,, \label{eq:Hertz}
\end{equation}
where $x$ is the overlap between the two particles and $A$ is the material parameter (see also Eq.~\eqref{material} below). A Hertzian interaction between a pair of particles occurs only when they are in contact, so each particle is affected directly only by its nearest neighbors and experiences a force from a neighbor only when it overlaps with it.  This yields the bracket
\begin{equation}
[x]_+ = \left\{
	\begin{array}{lcc}
		x\,, & \text{if} & x>0\,,\\
		0\,, & \text{if} & x\leq0
	\end{array}\right. 
\label{tensionless}
\end{equation}
in Eq.~\eqref{eq:Hertz}. The original derivation of the contact law \eqref{eq:Hertz} is in Hertz's seminal work \cite{Hertz}, and \cite{Johnson} has a more modern version of it. The full derivation described in \cite{Hertz,Johnson} is somewhat cumbersome, and Sen et al. gave a brief ``back-of-the-envelope" derivation in \cite{sen08}.  The nonlinear exponent $p$ depends on the geometry of the contact between particles. For example, $p=3/2$ for spherical particles. Cylindrical particles whose axes lie in the same plane but are not parallel also have an exponent of $p=3/2$. In this case, the material parameter $A$ depends on axis orientation. Other particle geometries can lead to exponents other than 3/2. For example, hollow spheres have interaction laws that vary as a function of shell thickness \cite{Johnson}. 

Assuming that the only relevant force is due to elastic contact between particles (see Fig.~\ref{fig:Hertz} for an illustration),
the 1D equations of motion for a finite-length granular chain (i.e., 1D granular crystal) are
\begin{equation}
	\ddot{u}_n = \frac{A_n}{m_n}[\delta_{0,n}+u_{n-1}-u_n]_+^{p}-
	\frac{A_{n+1}}{m_n}[\delta_{0,n+1}+u_n-u_{n+1}]_+^{p}\,,
\label{eq:model}
\end{equation}
where $u_n$ is the displacement of the $n$th particle (with $n\in \{1,2,{\ldots},N\}$, where $N$ is the number of particles) measured from its equilibrium position in an initially compressed chain, $m_n$ is the mass of the $n$th particle, and 
$\delta_n = ({F_0}/{A_n})^{1/p}$
is a static displacement for each particle that arises from the static load $F_0 =
\text{const}$. (See Fig.~\ref{fig:Hertz} for an illustration.)
For spherical particles, the exponent is $p=3/2$ and the parameter $A_n$, which reflects the material and the geometry of the chain's particles, has the form
\begin{equation} \label{material}
	A_{n} = \frac{4 E_{n}E_{n+1}\sqrt{\frac{R_{n}R_{n+1}}{\left(R_{n}+R_{n+1}\right)}}}%
{3E_{n+1}\left(1-\nu_{n}^2\right) + 3 E_{n}\left(1-\nu_{n+1}^2\right)  }\,,
\end{equation}
where $E_{n}$ is the elastic (Young) modulus of the $n$th particle, $\nu_{n}$ is its Poisson ratio, and $R_n$ is its radius. Important special cases of Eq.~\eqref{eq:model} include monoatomic (i.e., ``monomer'') chains (in which all particles are identical, so $A_n=A$, $m_n = m$, and $\delta_{0,n} = \delta_{0}$),  period-2 diatomic chains, and chains with impurities (e.g., a ``host'' monomer chain with a 
small number of ``defect'' particles of a different type). The simplest type of period-2 diatomic chain (which is sometimes simply called a ``dimer'' as a shorthand\footnote{The term ``dimer" can alternatively be defined as any chain that consists of two types of particles but are not necessarily arranged with a spatial period of 2. In this article, we assume that the spatial period of the dimer is 2 unless we specify otherwise.}) consists of alternating particles of two types, so $A_n=A$, $\delta_{0,n} = \delta_{0}$, and the mass is $m_n = m_0$ for even $n$ and $m_n = m_1$ for odd $n$. In such a chain, the mass ratio $m_1/m_0$ is the only additional parameter beyond the monomer case. We use the term ``$M$-mer'' for period-$M$ chains with $M > 2$.

We now present some basic analytical considerations for the equations of motion in Eq.~\eqref{eq:model}. It is convenient to work
in so-called ``strain variables'' $y_n = u_{n+1} - u_n$.  For a monomer chain, Eq.~\eqref{eq:model} becomes 
\begin{equation}
	\ddot{y}_n = \frac{A}{m}\left ( 2 [\delta_0 - y_{n}]_+^{3/2} - [\delta_0 - y_{n-1}]_+^{3/2}
	-[\delta_0 -y_{n+1}]_+^{3/2} \right)\,.
\label{strain0}
\end{equation}
Introducing the scaling $ t \rightarrow t \sqrt{\frac{A}{m}}$ allows us to write Eq.~\eqref{strain0} as
\begin{equation}
	\ddot{y}_n = \left ( 2 [\delta_0 - y_{n}]_+^{3/2} - [\delta_0 - y_{n-1}]_+^{3/2}
	-[\delta_0 -y_{n+1}]_+^{3/2} \right)\,.
\label{strain}
\end{equation}
If $\delta_0 \neq 0$, one can also normalize the precompression by rescaling the amplitude $y_n$ \cite{Nester2001}.
For an infinite lattice (i.e., $n\in\Z$), Eq.~\eqref{eq:model} has the Hamiltonian
\begin{equation*}
	H = \sum_{n\in\Z} \frac{1}{2} m_n \dot{u}_n^2 + V_n(u_{n+1} - u_n)\,,
\end{equation*}	
where the potential function is
\begin{equation}
	V_n(y_n) =  \frac{2}{5} A_{n+1}[ \delta_{0,n} - y_n]^{5/2} + \phi_n\,,  \quad  \mathrm{with} \quad \phi_n =  - \frac{2}{5} A_{n+1}[ \delta_{0,n} ]^{5/2} - A_{n+1}[ \delta_{0,n} ]^{3/2} y_n \,.      \label{pot}
\end{equation}
Note that $\phi_n$ in Eq.~\eqref{pot} implies that $V_n'(0) = 0$ and $V_n''(0)> 0$, which is necessary to ensure that the classical ground state (for which $u_n = \dot{u}_n=0$) is a minimum of the energy $H$ \cite{MacKay99}.

Equation~\eqref{eq:model} is the simplest model for the description of a granular chain, and it neglects many
features that lead to disparities between its solutions and experimental observations. One potentially important
feature is particle rotation \cite{merkel,cabaret,magneto}, and another is that the contact points of particles cannot be aligned perfectly in experiments because of clearance between the particles and support rods \cite{Lin}. This is likely to result in dynamic buckling of a granular chain when particles exhibit high-amplitude motion. In 1D settings, effects from the above two features are negligible in a variety of experimental configurations, but the role of rotation is considerably more important in higher-dimensional settings \cite{merkel}.

Additionally, as we explained above, the model of Eq.~(\ref{eq:model}) is Hamiltonian
in nature. This is a useful idealization for understanding the principal wave phenomenology in granular chains, but realistic granular crystals have dissipative effects (e.g., from wave attenuation) that can significantly modify the coherent-structure dynamics from those predicted by a Hamiltonian description like Eq.~(\ref{eq:model}). A customary approach is to incorporate a dashpot form of dissipation, $F_d^{(1)}=-\dot{u}_n m_n /\tau_n$ (where $\tau_n$ is a characteristic
time scale), to describe experimental observations qualitatively and
even quantitatively in some cases (upon parameter fitting)~\cite{Nature11,hooge12}. 
Another recent proposal~\cite{rosas2007,rosas2008} is to explore a form
of dissipation that involves a discrete Laplacian in the velocities.\footnote{This idea is 
  reminiscent of Navier--Stokes damping; 
  see, e.g., the discussion in~\cite{peyrard} about the physical
interest in such a functional form for the creation of ``lattice
turbulence.''}
The specific form employed for the force was $F_d^{(2)}= m_n [ (\dot{u}_{n-1} - \dot{u}_n)/\tau_{n-1,n} - 
(\dot{u}_n-\dot{u}_{n+1})/\tau_{n,n+1}]$, which was considered in the context of a monomer chain, and it was assumed that 
 $\tau_{n-1,n}$ is constant along the lattice. Subsequent generalizations of this functional form include power laws of the relative velocities~\cite{khatriprl} and more complicated forms, such as~\cite{vergara} 
\begin{equation*}
	F_d^{(3)} \propto m_n \left(\sqrt{u_{n-1} - u_n} (\dot{u}_{n-1} - \dot{u}_n) - \sqrt{u_{n} - u_{n+1}} (\dot{u}_{n} - \dot{u}_{n+1})\right)\,,
\end{equation*}	
which is inspired by viscoelasticity theory and involves both displacements and velocities.
Additionally, see \cite{ortiz2012} for an attempt to use finite-element modeling to derive a form for dissipative effects in granular crystals. Despite the multitude of proposed models for incorporating dissipation, presently there is no universally accepted 
functional form to describe dissipative effects in granular crystals. Features such as plastic deformation \cite{plastic1,plastic2} and damping due to rotation may also account for some of the wave
attenuation that has been observed in experiments~\cite{Yang_EXP_2014}. 
The correct way to model such effects in granular crystals is an important open research problem.
 In the present review article, we mostly restrict our attention to Hamiltonian models of granular crystals (and especially Eq.~\eqref{eq:model}).

\section{Linear Regime and Dispersion}
\label{sec:linear}

\begin{figure} 
\centering
  \includegraphics[width = .8 \linewidth]{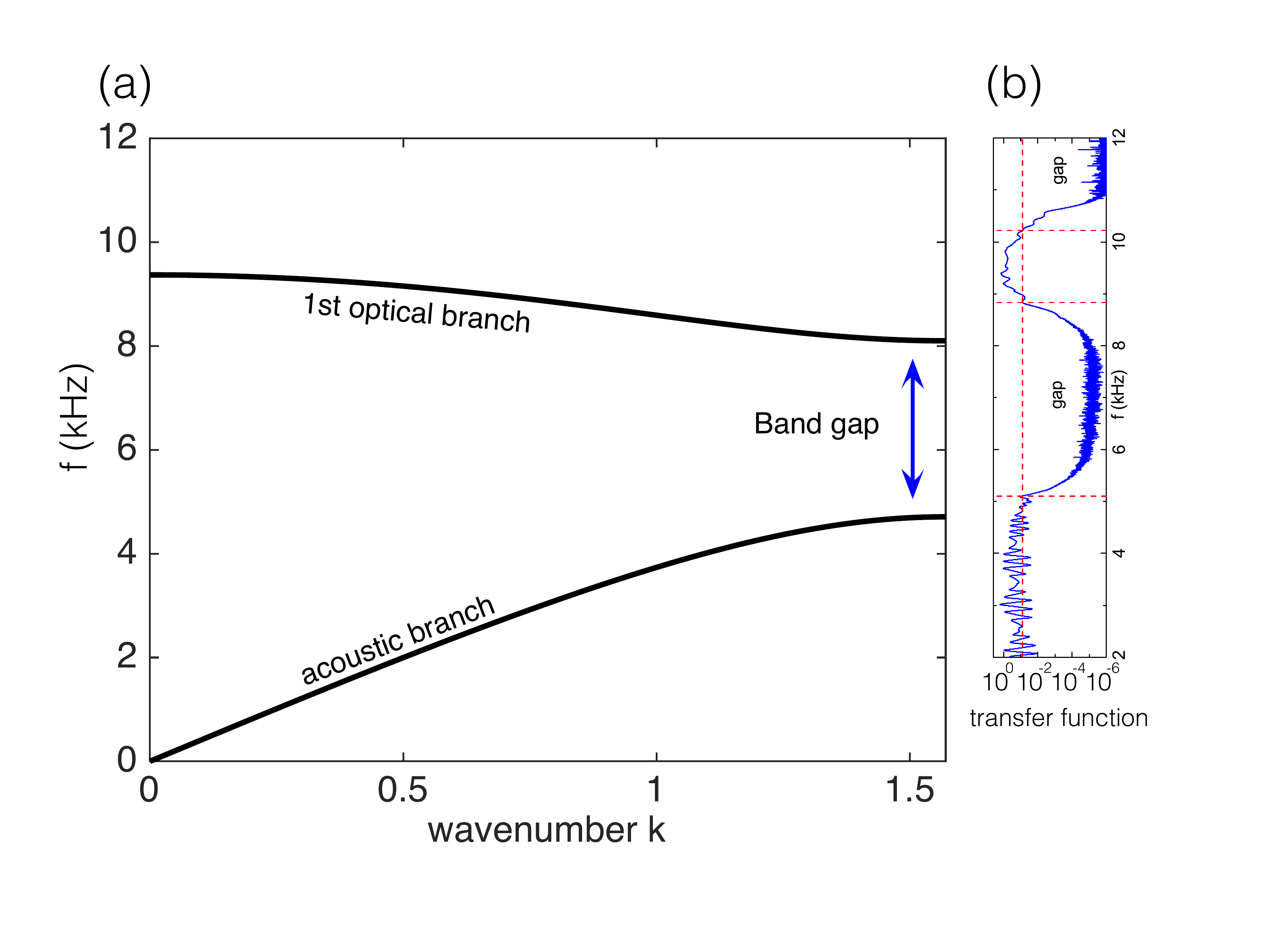} 
\caption{Pass bands (acoustic and optical) and band gap for the diatomic granular chain considered in \cite{Boechler2010}.
 \textbf{(a)} Analytical expression derived from the corresponding dispersion relations. \textbf{(b)} Experimental 
 transfer function inferred when sending ``white noise'' from one side through the 
system and then retrieving (on the other side) only the frequencies that belong
to the pass bands. [Panel (b) is adapted with permission from Ref.~\cite{Boechler2010}. Copyrighted (2010) by the American Physical Society.]
}
\label{fig:dispersion}
\end{figure}

The nature of the equations of motion \eqref{eq:model} is reminiscent of a Fermi--Pasta--Ulam (FPU) oscillator chain \cite{fpupop,FPUreview,FPU55}. The connection is especially evident in the weakly nonlinear regime, in which the dynamic strains are small compared with the static precompression:
\begin{equation}
	\frac{|u_{n}-u_{n+1}|}{\delta_{0,n}}\ll 1 \,.
\label{lin_limit}
\end{equation}
One can then Taylor expand the potential function to obtain
\begin{equation}
	 V_n'(y_n) \approx K_{2,n} y_n + K_{3,n} y_n^2 + K_{4,n} y_n^3\,, \label{eq:taylor}
\end{equation}
where
\begin{equation} \label{GCcoeff}
 	K_{2,n} = \frac{3}{2}A_{n+1}\delta_{0,n}^{1/2}\,, \quad K_{3,n} = -A_{n+1}\frac{3}{8} 
\delta_{0,n}^{-1/2}\,, \quad K_{4,n} = -A_{n+1}\frac{3}{48}\delta_{0,n}^{-3/2}\,.
\end{equation}
The function $V_n(y)$ in Eq.~\eqref{GCcoeff} is the classical FPU potential (also called the ``$K_2$--$K_3$--$K_4$ potential'' or the ``$\alpha$--$\beta$ potential'') \cite{FPUreview}. The corresponding linear problem,
\begin{equation} \label{eq:linear}
	\ddot{u}_n = \frac{   K_{2,n}    }{m_n} ( u_{n-1} - u_n) -     \frac{K_{2,n+1}}{m_n} (u_{n}- u_{n+1})\,,
\end{equation}
is a prototypical example of a coupled spring--mass system that serves as a textbook model for vibration modes \cite{Kittel}.
For a monomer chain, Eq.~\eqref{eq:linear} has plane wave solutions,
\begin{equation*}
	u_n(t) =  e^{i(k n + \omega t)}\,,  \qquad k\in[0,\pi]\,,
\end{equation*}	 
where the angular frequency $\omega$ and the wavenumber $k$ are related through the dispersion relation
\begin{equation}\label{eq:disp}
	[\omega(k)]^2 =  \frac{4K_2}{m}\sin^{2}(k/2) \,.
\end{equation}
Equation~\eqref{eq:disp} implies that the largest possible angular frequency
of a plane wave is $\omega_0 = 2 \sqrt{K_2/m}$ (the so-called ``cutoff point''). It thereby highlights the role of the lattice dynamical system \eqref{eq:linear} 
as a filter that allows the transmission of frequencies $\omega < \omega_0$ but prohibits the transmission of frequencies $\omega> \omega_0$. The sound speed $c_s$, which indicates the maximum speed of a linear wave, is defined as the phase velocity $\omega(k)/k$ in the limit $k \rightarrow 0$. In this case, $c_s = \sqrt{K_2/m}$. The single (positive) branch of the dispersion curve represented by Eq.~\eqref{eq:disp} is called the ``acoustic band'' of the linear spectrum. If the particles in a granular chain are arranged in a periodic manner with period $M\in\Z^+$ (e.g., $M=2$ for a dimer), then the linear
problem \eqref{eq:linear} is solved by Bloch waves 
\begin{equation} \label{eq:bloch}
	u_n(t) =  f_ne^{i(k n + \omega t)},  \qquad f_n = f_{n+M} \in \C, \quad k\in[0,\pi/M]\,.
\end{equation}
One can compute the dispersion relation by solving the $M\times M$ matrix eigenvalue problem obtained by substituting Eq.~\eqref{eq:bloch} into Eq.~\eqref{eq:linear}. This results in $M$ eigenvector--eigenvalue pairs $(\bf{f},-\omega^2)$, where ${\bf f} = (f_1, f_2 ,\ldots , f_M)^T$ \cite{Kittel}. We use notation so that eigenvalues are ordered as $-\omega_1^2(k) > -\omega_2^2(k) > \ldots > -\omega_M^2(k)$. 
The branch corresponding to $-\omega_1^2(k)$ is called the ``acoustic band,'' and $-\omega_j^2(k)$ is called the $j-1$th ``optical band'' (where $j>1$). See Fig.~\ref{fig:dispersion} for an example of a dispersion relation in a diatomic granular chain. If the chain has finite length or is not arranged in a periodic manner, one can obtain the angular frequencies of the linear modes (which are not necessarily sinusoidal in space) numerically by solving the $N\times N$ eigenvalue problem that results by substituting the ansatz $u_n(t) = \mathbf{v} e^{i\omega t}$, with $\mathbf{v} = (v_1,v_2, \ldots, v_N)^T$, into Eq.~\eqref{eq:linear}, where $N$ is the length (i.e., number of particles) of the chain.
In some special cases, one can approximate these eigenvalues analytically. For example, this is possible for a granular chain with a small number of impurities \cite{Craster2013,Defect_Man}. See \cite{Defect_Job} for a relevant study for strongly nonlinear chains, and see \cite{Theocharis2009} for a relevant study of weakly nonlinear chains. In a finite-length granular chain, the boundary conditions and length affect the mode shapes and introduce new modes, such as gap modes and edge modes (i.e., surface modes)\cite{Wallis57}.

Having reviewed the dynamics of linear granular chains, we now turn to the effects of nonlinearity in the next several sections. As we will see, nonlinearity has a significant impact on the formation, propagation, and aggregate dynamics of waves. In our discussions, it will nevertheless be helpful to keep the linear dynamics in mind.

\section{Traveling Solitary Waves } \label{sec:TW}

\subsection{Continuum Modeling and the Nesterenko Solitary Wave} \label{sec:qc}

\begin{figure} 
    \centerline{
    \includegraphics[width=\linewidth]{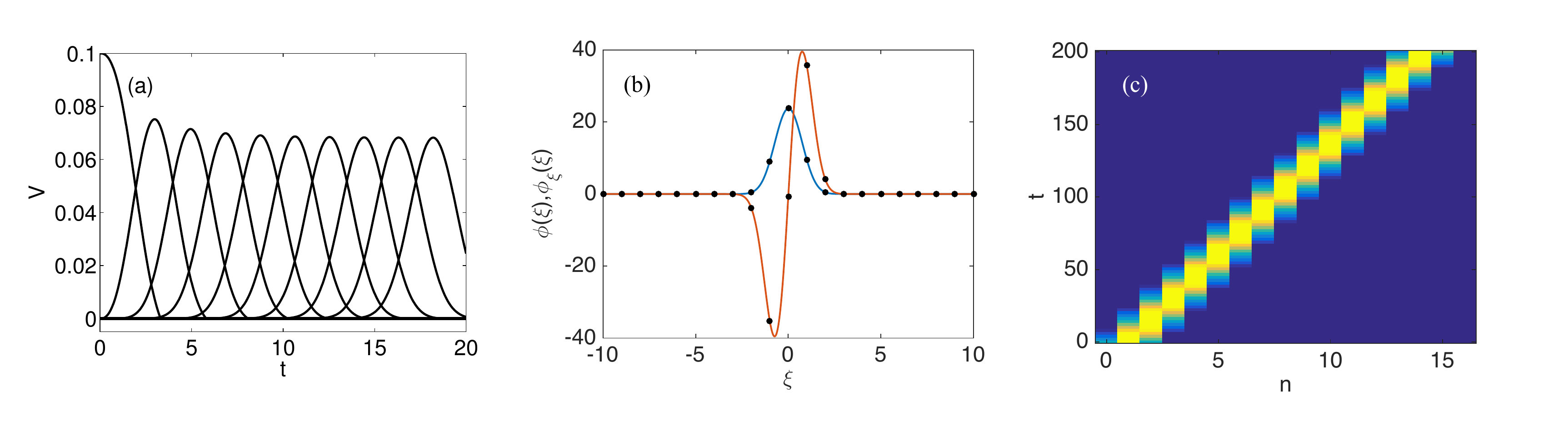} }
\caption{ \textbf{(a)} For a monomer granular chain, a compression wave travels through the chain after it is struck at one end. After the wave has traveled a few lattice sites, it appears as though a constant shape profile forms and propagates through the chain. Each bell-shaped curve corresponds to the velocity profile of a single particle. \textbf{(b)} Solution of Eq.~\eqref{adv-del} (solid curves), which represents a genuine traveling solitary wave. The markers are the corresponding location on the lattice.
\textbf{(c)} Contour plot of the evolution of Eq.~\eqref{eq:model} with initial data given by the solitary wave in panel (b).
}
\label{fig:TW}
\end{figure}

In 1985, Lazaradi and Nesterenko reported a new type of solitary wave \cite{Lazaradi} (see \cite{soliton-scholarped} for a review of solitary waves) that they found by striking one side of an uncompressed monoatomic granular chain with another particle. The resulting waveform was localized like a traditional solitary wave \cite{Zakharov}, but its tails exhibited much faster decay, and it was pondered whether the waves might even have
compact support~\cite{Nester2001}.\footnote{A solitary wave with compact support is called a ``compacton" \cite{Rosenau93}.} It was a new type of solitary wave (see Fig. \ref{fig:TW}). 
The experiments of Lazaradi and Nesterenko \cite{Lazaradi} were performed in the absence of static precompression. In this case, Eq.~\eqref{eq:model} is purely nonlinear, and it is very challenging to study it analytically because the nonlinearity is not smooth (as the quantity $V''(0)$ is undefined in the absence of precompression). One of the first analytical results on a purely nonlinear granular chain was due to Nesterenko \cite{Nester2001}, who derived an explicitly-solvable partial differential equation (PDE) as an approximation to Eq.~\eqref{eq:model}. This  follows a standard procedure in discrete models by using a long-wavelength (i.e., near-continuum) limit in which one expects that the spatial extent of the wave is much shorter than the spacing of the underlying lattice.
A standard mechanism to transfer from the original discrete framework
to this continuum limit is to introduce a small parameter $\epsilon$ and to define a new
variable $U(X,T)  \approx u_n(t)$ that approximates a solution to Eq.~\eqref{eq:model}, where 
the independent variables $X,T$ generally depend on $\epsilon$. 
In the remainder of our article, we use the notation $U(X,T)$ to represent a continuum approximation of the displacement $u_n(t)$, and we use $Y(X,T)$  to represent a continuum approximation of the strain $y_n(t)$. We consider different scalings throughout the review but we keep the same notation for simplicity. In our first example, we define $X = \epsilon n$ and $T= \epsilon t$.
Differences become partial derivatives through a Taylor expansion of the new variable $U$. We thus write
\begin{equation*}
	U(\epsilon(n \pm 1), T)  = U(X,T) \pm \pa_X U(X,T) \epsilon + \pa^2_X U(X,T) \frac{\epsilon^2}{2} \pm \pa^3_X U(X,T) \frac{\epsilon^3}{3!} + \pa^4_X U(X,T) \frac{\epsilon^4}{4!} + \cdots \,,
\end{equation*}
where $\partial_X^j$ denotes the $j$-th partial derivative with respect to the variable $X$. Because $\epsilon$ is supposed to be small, the new variable $U(X, T)$ has a wavelength that is long relative to that of the underlying lattice. Using the scaling $X = \epsilon n$, $T= \epsilon t$, one can derive a continuum model from Eq.~\eqref{eq:model} that is equivalent to the one Nesterenko derived: 
\begin{align}
	\pa_T^2 U= -\pa_X \left[  (-\pa_X U)^{p} + \frac{\epsilon^2}{12} \left( \pa_X^2 (-\pa_X U)^{p}  - \frac{p(p-1)}{2} (-\pa_X U)^{p-2}   \pa_X^2 U^2\right)   \right]\,,
\label{nester}
\end{align}
where we have assumed that the parameters are normalized (i.e., $A=m=1$). See \cite[Chap 1.3]{Nester2001} for the original derivation, and see \cite{mason2} for a detailed version of the derivation in the case of a dimer. 
The derivation of Eq.~\eqref{nester} assumes a long wavelength (i.e., $\epsilon \rightarrow 0$), although traditionally scholars have set $\epsilon = 1$. Consequently, one can construe Eq.~\eqref{nester} with $\epsilon =1 $ as a ``quasicontinuum" model, as it does not possess a small parameter, so it is not a
self-consistent asymptotic procedure. This is one of the limitations of Eq.~\eqref{nester}, as we will discuss in Sec.~\ref{sec:more_qc}.  
Let's press on with Eq.~\eqref{nester} despite its limitations and seek traveling wave solutions. We define the coordinate $\xi = X-cT$ and the new strain variable $r(\xi) = U_\xi(\xi)$. Substituting the traveling wave ansatz $r(\xi)$ into Eq.~\eqref{nester} 
leads to an ordinary differentially equation (ODE) that can be solved explicitly \cite[Chp 1.3]{Nester2001}. Among the solutions of this ODE, one of particular interest is (what has come to be known as) the ``Nesterenko solitary wave solution'' (which is also sometimes called the ``Nesterenko soliton'' or ``Nesterenko compacton'') \cite{Nester2001,pikovsky},
\begin{equation} \label{nesterwave}
	r(\xi) = \left\{
		\begin{array}{lr}
			\left(\frac{2}{1+p}\right)^{\frac{1}{1-p}}  |c|^{\frac{2}{p-1}} \cos^{\frac{2}{p-1}}
			\left( \xi \sqrt{6\frac{(p-1)^2}{p(p+1)}}    \right)\,,  &   |\xi| < \pi  \sqrt{\frac{p(p+1)}{24(p-1)^2}} \,,  \\
			0\,, &  \mbox{otherwise}\,.
\end{array} \right.
\end{equation}
Note that if one starts with the strain-variable formulation Eq.~\eqref{strain} and performs the same derivation
for a new continuum approximation $Y(X,T)$, then $Y(\xi) \neq r(\xi)$. See the derivation and discussion in \cite{pikovsky}.

In the approximation \eqref{nesterwave}, it is apparent that the amplitude of the
wave scales with its speed $c$ (a nonlinear analog of a dispersion relation) and that the wave has a nonzero value only for a finite number of lattice sites (roughly 5 particles for a Hertzian interaction). 
Specifically, this suggests that the wave is {\it genuinely} compact and
thus takes the form of a compacton.
However, as we show in Sec.~\ref{sec:genuineTW}, this is not the case, although the wave does possess very rapidly decaying (doubly exponential) tails.

In simulations and experiments, one can generate a traveling wave in a granular chain by impacting one boundary of the chain (which is at rest) with some initial velocity $v_i$. Using scaling arguments on the quasicontinuum model \eqref{nester}, one can show that $c \propto v_i^{1/5}$ \cite[Sec. 1.4]{Nester2001} for Hertzian interactions. The analytical approximations that we reviewed in this subsection are consistent with experimental and numerical findings \cite{Lazaradi,Coste,Coste1999}.
Such results are now considered classical, and numerous studies on traveling waves in granular chains have followed them and built upon them. See the review article \cite{sen08}, which is dedicated to solitary wave solutions of granular chains. In the remainder of this article, we will emphasize results that appeared after \cite{sen08}.

\subsection{Genuine Traveling Solitary Waves} 
\label{sec:genuineTW}

The experiments, numerical simulations, and informal asymptotics that we discussed in Sec.~\ref{sec:qc} all suggest the existence
of stable traveling solitary wave solutions in a homogeneous granular chain. In this subsection, we consider the fundamental mathematical issue of the existence of \emph{genuine} traveling wave solutions.  Genuine traveling waves are solutions that travel through a granular chain without any disturbance to the original wave shape. To identify genuine traveling wave solutions, it is convenient to work directly in the strain-variable formulation of Eq.~\eqref{strain}. Substituting the ansatz $y_n(t) = \Phi(n - ct,t) = \Phi(\xi,t)$ into Eq.~\eqref{strain} leads to
\begin{equation}\label{step_2}
 	\pa_t^2{\Phi}(\xi,t)=-c^{2} \pa_\xi^2{\Phi}(\xi,t)+2c%
 \pa_{\xi \,t }{\Phi}(\xi,t)+ \Big\lbrace\left[\delta_{0}+ {\Phi}(\xi-1,t)\right]_{+}^{3/2}-%
2\left[\delta_{0}+ {\Phi}(\xi,t)\right]_{+}^{3/2}+\left[\delta_{0}+ {\Phi}(\xi+1,t)\right]_{+}^{3/2}\Big\rbrace\,.
\end{equation}
Traveling waves of Eq.~\eqref{eq:model} correspond to stationary (i.e., time-independent) solutions $\Phi(\xi,t) = \phi(\xi)$ of Eq.~\eqref{step_2}. Such solutions satisfy
\begin{equation} \label{adv-del}
	0 = - c^{2}\pa_\xi^2\phi + \lbrace \left[\delta_{0}+\phi(\xi-1)\right]_{+}^{3/2}%
		-2\left[\delta_{0}+\phi(\xi)\right]_{+}^{3/2}
	            +\left[\delta_{0}+\phi(\xi+1)\right]_{+}^{3/2}\rbrace\,.
\end{equation}
If the solution also has the property that $\lim_{\xi \rightarrow \pm \infty} \phi(\xi) = 0$, it is called a ``solitary wave.''
The existence of such solutions follows from the general result of Friesecke and Wattis \cite{Wattis}, which
holds for traveling wave solutions of general FPU lattice equations. In \cite{MacKay99}, this result was applied to Eq.~\eqref{strain}. Although this existence result is exact (no approximations were made), these proofs are not constructive, and thus analytical approximations like \eqref{nesterwave} and numerical computations are very useful. Chatterjee argued heuristically in \cite{Chatterjee} that the tails of this genuine traveling wave solution are doubly exponential rather than truly compact (as suggested by the continuum approximation of Eq.~\eqref{nesterwave}) when there is no precompression. This fact was proved subsequently in \cite{pego1} by formulating the advance--delay equation \eqref{adv-del} (with $\delta_0 = 0$) as a fixed-point problem: Using the transform $\phi(\xi)=\int_{-\infty}^{\infty} \hat{\phi}(k) e^{i k \xi} dk$ leads from Eq.~(\ref{adv-del}) to
\begin{equation} \label{fft}
  \hat{\phi}(k) = \frac{1}{c^2} {\rm sinc}^2\left(\frac{k}{2}\right)\widehat{\phi^{3/2}}\,, 
\end{equation}
where $\, \widehat{} \,$ denotes the (wavenumber-dependent) Fourier transform of the spatial variable. Invoking the convolution theorem yields
\begin{align} \label{fp}
	\phi(\xi) = K \ast \phi^{3/2} = \int_{-\infty}^{\infty} \Lambda(\xi-y) \phi^{3/2}(y) dy\,,
\end{align}
where $\Lambda(\xi)= (1/c^2) \max\{1-|\xi|,0\}$. Without loss of generality (given the scaling properties of Eq.~\eqref{strain}),
one can assume that $c=1$.
Using $z=\xi-y$ and changing variables in~\eqref{fp} then yields
\begin{align} \label{eqn23}
	\phi(\xi+1) =\int_{-1}^1 \Lambda(z) \phi^{3/2}(\xi+1-z) dz 
	\leq 
	\phi^{3/2}(\xi) \Rightarrow \phi(\xi+n) \leq \phi(\xi)^{(\frac{3}{2})^n}\,,
\end{align}
where we used only the fact that ${\rm sup}_{y \in [-1,1]} \phi^{3/2}(\xi+1-y)=\phi^{3/2}(\xi)$
for a positive, monotonically decreasing traveling wave
(and the normalization $\int_{-1}^1 \Lambda(z) dz=1$).
This establishes the doubly-exponential decay for traveling solitary waves in granular chains with a Hertzian potential $p = 3/2$, and it shows for general exponents $p$ that $\phi_n \sim \phi_0^{p^n}$ as $n \rightarrow \infty$.
Equation~\eqref{fp} also provides an efficient 
numerical algorithm for the computation of such waves by setting up a fixed-point iteration scheme (see \cite[Sec. 4]{pego1}). 
Reference \cite{Stefanov} used a reformulation of Eq.~\eqref{fp} to prove the existence of bell-shaped, monotonically decaying traveling waves for $\xi > 0$ by constructing an energy functional for which the constrained minimization problem over bell-shaped entries has a solution. The proof was generalized in~\cite{Stefanovb} for granular chains with nonzero
precompression. In the latter scenario, the precompression absorbs most of
the decay in the estimate of Eq.~\eqref{eqn23}, leaving 
$\phi(\xi+1)$ bounded above by $\phi(\xi)$ multiplied by a suitable prefactor. Decay in this case is thus exponential, as observed both numerically and experimentally \cite{Nester2001}. The associated waves are strongly reminiscent of the supersonic waves of standard FPU chains~\cite{Stefanovb}.
Moreover, the speed of sound is related to the precompression: $c_s \propto \sqrt{p} \delta_0^{\frac{p-1}{2}}$, in line
with the characterization of the $\delta_0=0$ case as a
``sonic vacuum''~\cite{Nester2001}. 
We will further discuss granular chains with precompression in Sec.~\ref{sec:preTW}. 

One can study the stability of stationary solutions $\phi^0$ of Eq.~\eqref{adv-del} by examining the fate of small perturbations around them. More concretely, one substitutes the linearization ansatz $\Phi(\xi,t) = \phi^0 + \varepsilon a(\xi) e^{\lambda t} $ into Eq.~\eqref{step_2} to obtain the eigenvalue problem
\begin{equation} \label{eig1} 
	\small
	   \lambda^2 a(\xi) =  -  c^{2} \pa_{\xi}^2{a}(\xi)+2 \lambda c%
	 \pa_\xi{a}(\xi)+\frac{3}{2} \Big\lbrace\left[\delta_{0}+ {\phi}^0(\xi-1)\right]_{+}^{1/2} a(\xi-1) -%
	2\left[\delta_{0}+ {\phi}^0(\xi)\right]_{+}^{1/2}a(\xi)+\left[\delta_{0}+ {\phi}^0(\xi+1)\right]_{+}^{1/2}a(\xi+1)\Big\rbrace\,.
\end{equation}
A similar formulation was used to study the stability of traveling kink solutions of the sine-Gordon equation \cite{Jones}. The stability properties of genuine solitary wave solutions are unknown from a mathematically rigorous perspective, although both direct numerical simulations and experiments overwhelmingly suggest that they must be stable \cite{Nester2001,Stefanov2015}. Moreover, it is not even clear if the stability characteristics of stationary solutions of the advance--delay equation~\eqref{step_2} 
are inherited by the corresponding solutions of the discrete equation \eqref{strain}. More generally, this is an important question
about the stability of traveling waves on lattices, as the model of Eq.~(\ref{step_2})
contains spatial scales that are in some sense ``inaccessible'' to the original lattice
(i.e., ones that are below the scale of the lattice spacing). Consequently, one can, in principle,
envision a scenario in which Eq.~\eqref{step_2} includes instabilities
whose imprint cannot manifest in the original system in Eq.~\eqref{strain}. 

In the absence of rigorous mathematical results, numerical computations give another route to gather information on the stability of traveling waves in granular chains.  In this latter context, one requires a numerical approximation of Eq.~\eqref{adv-del}, which one can achieve by discretizing Eq.~\eqref{adv-del} in the variable $\xi$ and employing Newton iterations to approximate the solution \cite{Stefanov2015,origami,Duanmu}. Upon this discretization, Eq.~\eqref{eig1} becomes a standard matrix eigenvalue problem. One must be cautious, however, because the discretization can introduce spurious eigenvalues \cite{origami,spurious}. Indeed, the best choice of discretization for stability computations is also an open problem.

An important complementary approach to understanding the stability of the traveling wave solution of Eq.~\eqref{strain}
through analysis of Eq.~\eqref{step_2} that respects the underlying
discreteness of the lattice model is to consider traveling waves as periodic orbits imposed at the level of the original lattice~\cite{pegof3}. In other words, they are fixed points of the dynamical evolution process of travel by one site followed by 
an integer shift of the sites. This viewpoint was explored for traveling breathers
in~\cite{floria}. In our view, pursuing both of these approaches (discrete versus continuum), comparing their
spectra (see also the recent work of~\cite{Duanmu}), and appreciating
their similarities and differences is a crucial open area in the mathematical analysis of traveling
waves in discrete Hamiltonian systems.

\subsection{Other Continuum Models and Their Limitations} \label{sec:more_qc}

 \begin{figure} 
 \centering
\includegraphics[width=.9 \linewidth]{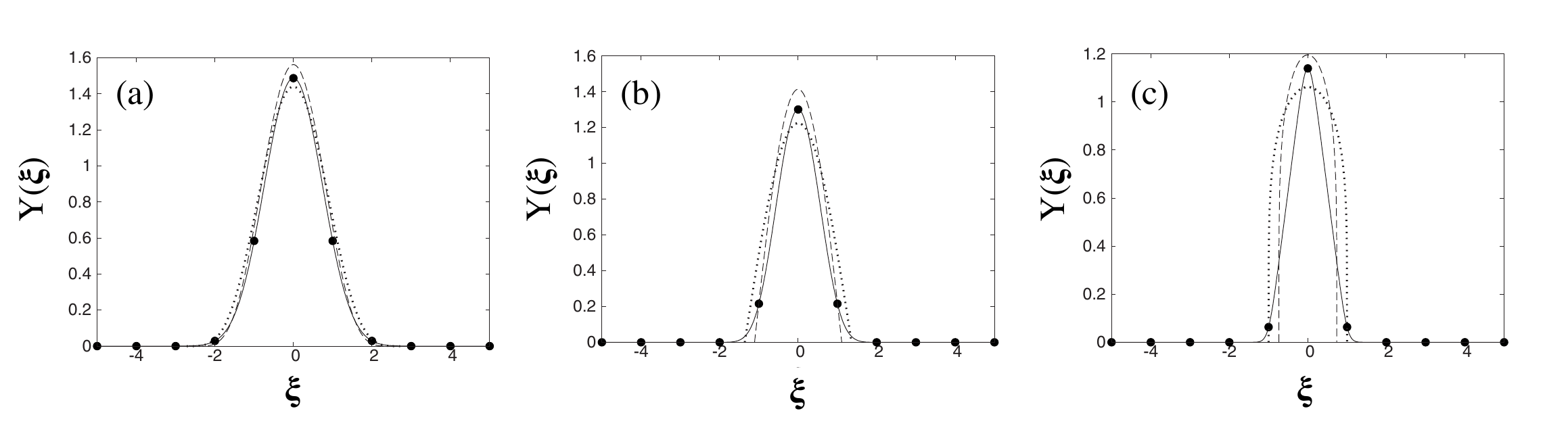}
\caption{Comparison of continuum approximations and genuine solitary wave solutions of Eq.~\eqref{strain}
for \textbf{(a)} $p=3/2$, \textbf{(b)} $p = 3$, and \textbf{(c)} $p = 11$. Markers show the wave on the lattice, dotted curves show the corresponding solutions of the continuum model \eqref{pik}, and dashed curves show Eq.~\eqref{nesterwave}. The approximation does poorer as the nonlinearity exponent $p$ increases. 
[The figure is adapted with permission from Ref.~\cite{pikovsky}. Copyrighted (2009) by the American Physical Society.]
}
\label{fig:QC_Compare}
\end{figure}

 Although the agreement between the quasicontinuum approximation \eqref{nester} and numerical computations seems satisfactory for the exponent $p=3/2$, the level of agreement becomes progressively less satisfactory as $p$ increases, as was discussed systematically in~\cite{pikovsky} (see Fig.~\ref{fig:QC_Compare}). Consequently, it is natural to ask if it is possible to derive the quasicontinuum model of Eq.~\eqref{eq:model} in a mathematically rigorous way---for example, by bounding the error of the approximation in a suitable norm in terms of some small parameter. Using the scaling $Y(\epsilon n, \epsilon t) = Y(X, T)$, one can derive the following continuum model from the strain-variable formulation \eqref{strain} of a granular chain \cite{pikovsky}:
\begin{align} \label{pik}
	\pa_{T}^2Y=\pa_X^2(Y^p) + \frac{\epsilon^2}{12} \pa_X^4(Y^p)\,.
\end{align}
Like Eq.~\eqref{nesterwave}, Eq.~\eqref{pik} also possesses solutions with compact support. Equation~\eqref{pik}
is a strongly nonlinear model reminiscent of the Boussinesq equation. Both \eqref{nester} and \eqref{pik} are pathological in the sense that they are ill-posed for $p=1$ because of large-wavenumber instabilities in the associated dispersion relation $\omega^2 = k^2 - \frac{\epsilon^2}{12} k^4$ \cite{badBoussinesq}. The fact that instabilities at large $k$ leads to ill-posedness of Eq.~\eqref{pik} has been studied both numerically and by demonstrating instability of the compact solutions \cite{pelingj}. Alternative paths~\cite{rosenau1,rosenau2} (see also \cite[1.4]{Nester2001})
for ``regularizing''  these equations are possible. For example, one can write
\begin{align} \label{ros}
	\pa_{T}^2 Y = \pa_X^2(Y^p) + \frac{\epsilon^2}{12} \pa_X^2\pa_{T}^2 (Y^p)
\end{align}
by inverting $1-\frac{\epsilon^2}{12} \partial_X^2$ in Eq.~(\ref{pik}) and Taylor expanding it on the left-hand side.
However, Eq.~\eqref{ros} also has issues, as now traveling waves acquire an 
exponential tail \cite{pelingj} and thus do not possess the proper doubly-exponential 
decay~\cite{pego1,Stefanov}. It is also interesting (and an open avenue) to consider 
a variant of Eq.~\eqref{pik} involving
Pad{\'e} approximations~\cite{rosenau1,rosenau2,titi} of the form
\begin{align}
	\pa_{T}^2 Y = \frac{\partial_X^2}{1-\frac{\epsilon^2}{12} \partial_X^2} Y^p\,.
\label{pade}
\end{align}
As an aside, we note that one can directly approximate the solitary traveling wave of Eq.~\eqref{eq:model} using a Pad{\'e} approximation by substituting the ansatz 
\begin{equation*}
	\phi(\xi) = \frac{1}{ \displaystyle \sum_{j} q_j \xi^j }
\end{equation*}	
 into a normalized variant of Eq.~\eqref{adv-del} and solving an algebraic system for the $q_j$ \cite{Yuli2010}. 
The continuum models above are somewhat unsatisfactory, as they require an arbitrary truncation in the Taylor expansion of the nonlinearity. A more mathematically sound approach is possible by defining $p = 1 + \epsilon$ and considering
an approximation in the $\epsilon \rightarrow 0$ limit (i.e., in the linear limit). One thereby derives the so-called ``log-KdV equation'' \cite{pelingj} 
\begin{align} \label{eqn28b}
	\pa_T Y+ \pa_X^3 Y + \partial_X (Y \log(|Y|))=0\,.
\end{align}
using the scaling $Y( \epsilon(n-t),\epsilon^3 t) = Y(X,T)$. Equation~(\ref{eqn28b}) possesses an exact Gaussian traveling wave
solution. The exact solitary wave solution of Eq.~\eqref{adv-del} and the compacton solution of \eqref{pik} both approach
this Gaussian profile in the $\epsilon \rightarrow 0$ limit for near-sonic wave speeds. However, the Gaussian traveling wave does not capture the doubly-exponentially decaying coherent structure as $p$ deviates from $1$ (and especially for $p=3/2$,
where it fails by about 35~\%). It is unclear if Eq.~\eqref{eqn28b} is well posed, as the nonlinear term diverges at $Y=0$.
Consequently, a full understanding of continuum modeling of a purely nonlinear granular chain remains a difficult problem that has only partially been addressed.

\subsection{Traveling Solitary Waves in Precompressed Granular Chains} \label{sec:preTW}

If a granular chain is compressed statically (such that $\delta_0 \neq 0$), its governing equations ~\eqref{eq:model} are linearizable and the resulting dynamics differ strikingly from the case without precompression. As we indicated previously (see Sec.~\ref{sec:genuineTW}), the supersonic waves in this case possess tails with an exponential decay (rather than the doubly-exponential decay of a purely nonlinear granular chain). This implies that the traveling wave solution of the purely nonlinear problem is a singular limit---rather than a regular limit---of the traveling wave solution as the precompression tends toward $0$. 

In the presence of precompression, the scaling $Y( \epsilon(n-t),\epsilon^3 t) = Y(X,T)$  leads to another log-KdV equation, similar to Eq.~\eqref{eqn28b}, that is well posed (because the nonlinear term is no longer divergent) \cite{Dumas}. 
This variant of the log-KdV equation is a useful model for general initial data, as opposed to Eq.~\eqref{eqn28b}, which is relevant only for special solutions. A rigorous justification of this log-KdV equation as an accurate continuum limit of a granular chain
with precompression (defined in Eq.~\eqref{strain}) was given in \cite{Dumas}.

Adding precompression also introduces a natural small parameter $y_n/\delta_{0,n} \ll 1$ (i.e., the strain amplitude $y_n$ relative to the precompression value $\delta_{0,n}$). For solutions with small strain $y_n$, the nonlinear response is weak, and one can describe the dynamics of a granular chain through its associated Taylor expansion. Continuum modeling of the FPU equation with a polynomial potential is well understood \cite{SW00,pegogf1}. For example, using the scaling $y_n = \eps^2 Y( \eps( n - ct),\eps^3 t) = \eps^2 Y(X,T) $ 
applied to a precompressed granular chain with the first three terms in its Taylor expansion (see Eq.~\eqref{eq:taylor}) 
leads to the Korteweg--de-Vries (KdV) equation~\cite{Zabusky} 
\begin{equation} \label{eq:kdv}
	2 c \pa_T Y =  \frac{c^2}{12} \pa_X^3Y + K_3 \pa_X(Y^2)\,,
\end{equation}
where one chooses $\delta_0$ so that $K_2=1$. The soliton solution of the KdV equation \eqref{eq:kdv} provides a reasonable approximation of the genuine traveling solitary wave solution of a precompressed granular chain
for velocities $c$ close to (but above) the sound speed $c_s$. solitary wave interactions have also been explored in this
context~\cite{Shen}. The KdV description has been studied rigorously (with error estimates) for a monomer FPU chain \cite{pegogf1,SW00}. In stark contrast to granular chains without precompression, there are rigorous results not only for the existence of traveling solitary waves but also in descriptions of both their linear and nonlinear stability~\cite{pegof2,pegof3,pegof4}.
Additionally, \cite{Bruckner10,Wright} rigorously derived the KdV equation for dimer and $M$-mer granular chains (see Sec.~\ref{beyondmon}).

\subsection{Traveling Structures in Configurations Other than Monomer Chains}\label{beyondmon}

 \begin{figure} 
 \centerline{
\includegraphics[width= \linewidth]{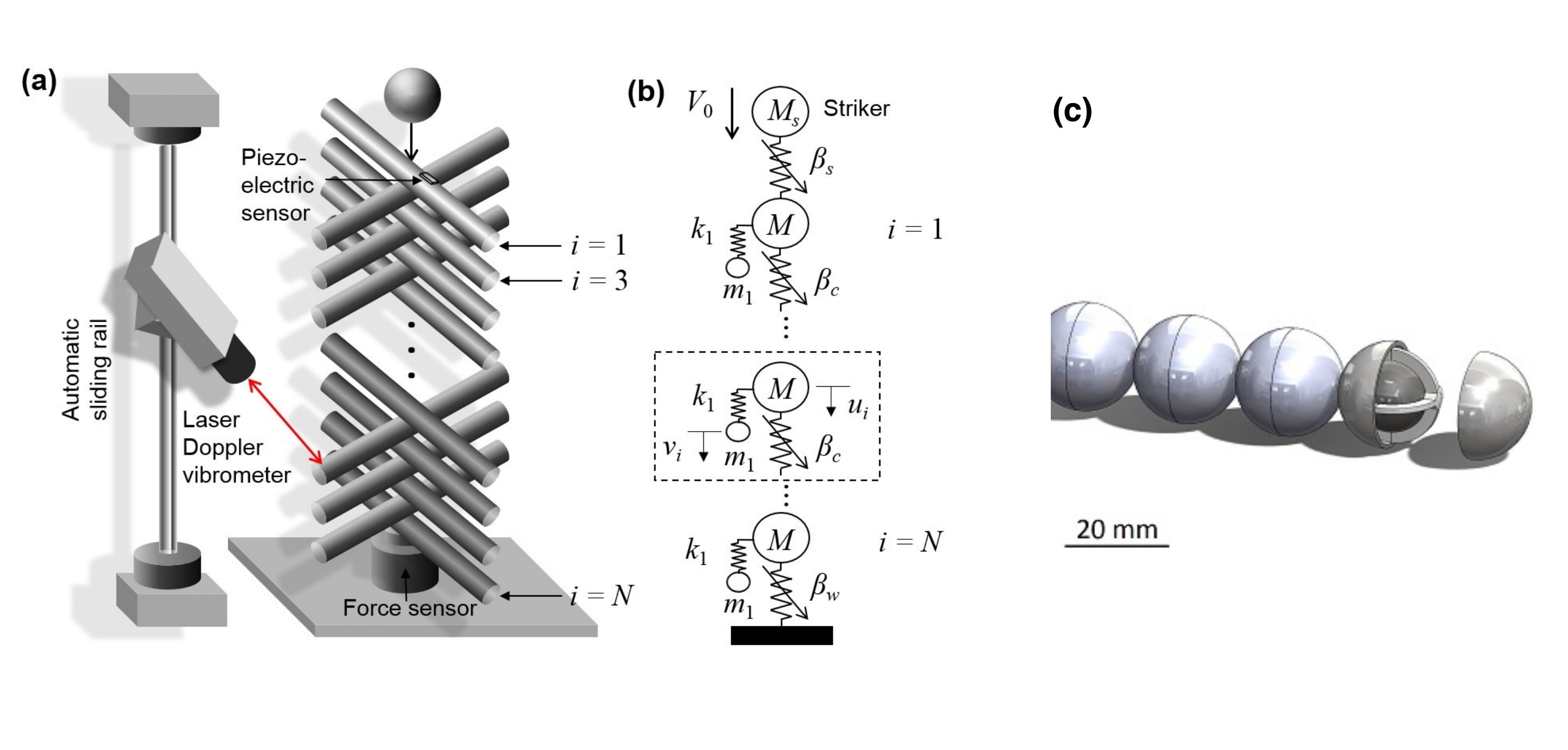} }
\caption{ \textbf{(a)} Example of a woodpile phononic crystal, which can be modeled as \textbf{(b)} a locally resonant
granular crystal. \textbf{(c)} A mass-in-mass granular chain.
[Panels (a,b) are adapted with permission from Ref.~\cite{jkprl}. Copyrighted (2015) by the American Physical Society.
Panel (c) was created by Luca Bonanomi and is used with permission from \cite{chiara_www}.]
}
\label{fig:local_res}
\end{figure}

Let's now discuss striking the end of a heterogeneous (i.e., nonuniform) granular chain. Generically, for heterogeneous chains, one does not observe the formation of a traveling structure with constant shape that propagates indefinitely (as in Fig.~\ref{fig:TW}).
For example, for a diatomic granular chain that consists of alternating light and heavy masses, the lighter mass particles vibrate between adjacent particles of heavier mass after the initial wave front has passed. This can result in the attenuation of a traveling wave. Hence, although early numerical and experimental (and heuristic analytical) studies for diatomic (and triatomic) granular chains 
suggested the possible existence of propagating structures~\cite{mason1,mason2},
a more systematic examination of the dynamics revealed 
an intriguing phenomenology that depends on the ratio of light to heavy masses~\cite{Jayaprakash1}.
This ratio is a canonical parameter of a diatomic chain, and pulse attenuation can be significant at special mass 
ratios (i.e., at ``resonances'')~\cite{Jayaprakash1,Jayaprakash2}, as has been studied both theoretically \cite{Jayaprakash2} and experimentally \cite{JKdimer,Nester_dimer}. Interestingly, for another set of special mass ratios (so-called ``anti-resonances'')~\cite{Jayaprakash1}, there seems to exist a genuine traveling wave 
structure, as the vibrating frequency of the lighter particles is in synchrony with the wave speed of the traveling structure, so energy is transferred completely from the lighter to the heavier particles. This latter
feature has also been confirmed experimentally~\cite{JKdimer}. (See \cite{Vainchtein} for the related case of a diatomic Toda lattice.) In contrast to the solitary waves of a monoatomic chain, the tails of a solitary wave in a dimer have small-amplitude oscillations that do not decay to $0$ \cite{Jayaprakash1,mason2}. It is an open problem if there are indeed special mass ratios in which the oscillating tails vanish (i.e., for which there exists a genuine solitary wave). This question has been partially answered in the related problem of the diatomic FPU problem with a polynomial potential \cite{faver}. There, it was shown rigorously that solitary waves exist near sonic speed and that they are the superposition of a periodic traveling wave (the tail) and
an exponentially localized function. However, \cite{faver} did not give a lower bound for the amplitude of the periodic traveling wave, so one cannot guarantee that this ``oscillating tail" is nonzero for all mass ratios. 
Their result applies to precompressed granular chain, but the methodology in \cite{faver} cannot be used for a purely nonlinear granular chain.
The exact numerical computation of solitary waves with nonvanishing tails in a granular crystal is an intriguing open problem, and it is also open to perform systematic numerical illustrations of this phenomenon, if it exists, in the strongly nonlinear limit.

A variety of similar features to the ones that we described in the paragraph above arise in ``locally resonant chains,'' in which each particle is connected to a vibrating (resonator) element. For a linear resonator, the equations of motion are
\begin{align} \label{eq:localres}
	m_n \ddot{u}_n &=  A_n [\delta_n+u_{n-1}-u_n]_+^{3/2}-
		A_{n+1}[\delta_{n+1}+u_n-u_{n+1}]_+^{3/2} - k_1(v_n - u_n)\,, \\
      \ddot{v}_n &= k_1(v_n - u_n) \,,
\end{align}
where $v_n$ is the displacement of the attached resonator (whose mass we have scaled to $1$ in Eq.~(\ref{eq:localres}).
Physical examples of locally resonant chains include ``mass-in-mass'' systems (in which each particle contains another vibrating particle) \cite{Daraio2015PRE,serra} and ``mass-with-mass'' systems (in which the resonators are external
rather than internal)~\cite{gantzounis,serra}. Another type of granular chain of a similar flavor that has been the subject of intense investigation recently are so-called ``woodpile'' phononic structures, which consist of a stacked array of cylinders (i.e., rods), whose bending modes are modeled as vibrating elements \cite{jkprl}.
solitary wave solutions with vanishing tails of locally resonant chains are similar to their diatomic counterparts. Generically, 
they have nonvanishing oscillating tails~\cite{Stefanov2015}, but their tails do vanish for some special parameter values. See \cite{Stefanov2015} for numerical computations and \cite{Stefanov2016} for a rigorous proof. The woodpile chains also have another interesting characteristic: because the resonators in this case are ``intrinsic'' (i.e., they consist of the vibrational
modes of the rods themselves), one can adjust the length (and other properties) of the cylinders to tune the associated internal vibration modes.
A scenario in which multiple such modes are involved
in the dynamics was broached in~\cite{jkprl}, and it seems to lead typically to decay of the wavelike excitations.

Traveling waves in heterogeneous granular chains with nonperiodic arrangements also exhibit interesting phenomena.
The effect of impurities on traveling waves in granular chains is well studied and results in reflected and transmitted waves \cite{sen08}. Intriguingly, in the presence of
multiple impurities (as well as precompression), the dynamics
can feature resonant interactions that enable (perfect) transmission
of small-amplitude waves through the impurities. This analog
of the well known Ramsauer--Townsend resonance in quantum physics was illustrated recently in~\cite{premason}.
Another type of heterogeneous chain is a tapered one, in which some feature (e.g., the mass) of the particles increases or decreases gradually along a chain (see, e.g., \cite{doney06}). One can also design particular heterogeneities to mitigate forces at the end of a chain \cite{fernando}.
While the quasicontinuum modeling described in Sec.~\ref{sec:qc}
cannot be applied to such situations, one can obtain
analytical approximations by employing the so-called ``binary collision approximation,'' in which at each time one considers the dynamics 
at only two lattice sites~\cite{linderberg}.

Another intriguing direction of intense recent interest involves
the interplay of structural disorder with the granular nonlinearity
and the lattice discreteness.
In the broad context of condensed-matter physics, disorder (e.g., in the form of randomness) causes strong localization---so-called ``Anderson localization''---that affects the diffusivity of waves in disordered media~\cite{anderson1957,Brandes2003}. Recent studies in the context of granular crystals have demonstrated computationally that such a phenomenon can be altered dramatically and can be even reversed (as superdiffusive transport can occur) by manipulating the nonlinearity strength in mechanical systems~\cite{Martinez_disorder_2016,Achilleos_2016} (see also, e.g., \cite{sokolow2007,Ponson:PRE2010}) and its potential ``competition'' with disorder. Such versatile dynamics associated with disorder and nonlinearity have only been studied very recently in granular crystals (and, more generally, they have not been studied much in lattice systems with inter-site interactions). There are numerous exciting directions to pursue in this area.
Experimental efforts to verify these predictions are presently
underway~\cite{Kim_disorder_2016}. 
A far more detailed understanding exists for lattice systems with on-site
nonlinearities~\cite{Flach_2015,Laptyeva_2014}. In the on-site case,
nonlinearity promotes localization (in the form of discrete breathers),
rather than enabling mobility (in the form of traveling waves,
as in FPU-type lattices). As a result, energy transport is only subdiffusive, whereas (as indicated above) superdiffusivity can occur for disordered granular crystals \cite{Martinez_disorder_2016,Achilleos_2016}. The study of disorder and nonlinear analogs of Anderson localization are important directions to pursue; they are accessible experimentally, and a systematic theoretical understanding of associated superdiffusive and subdiffusive transport should lead to important insights more generally on the interaction between disorder and nonlinearity.

\begin{figure} 
   \centering
  \includegraphics[width= \linewidth]{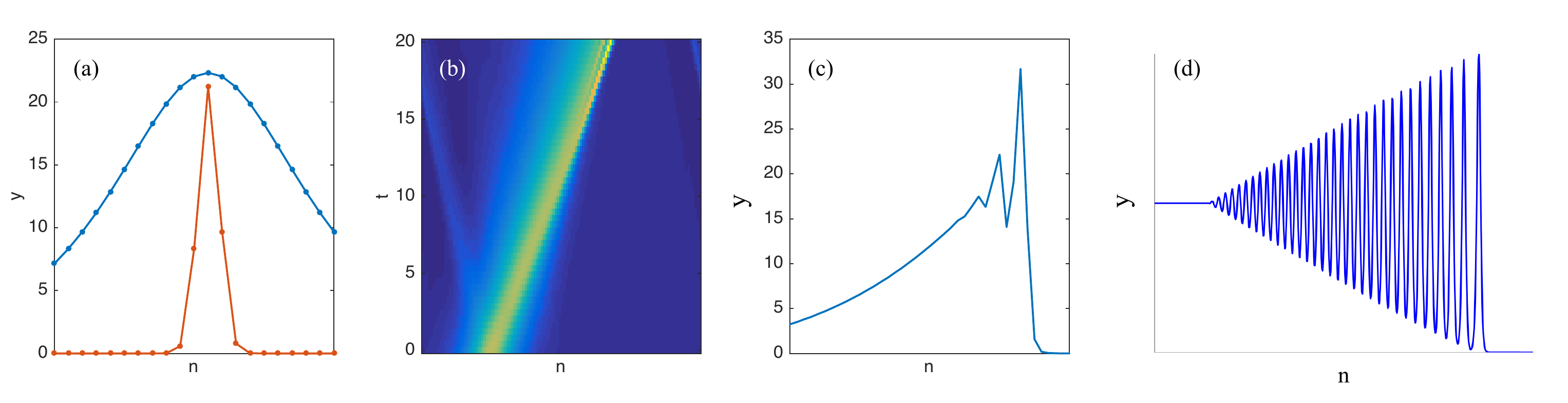} 
\caption{\textbf{(a)} Strain profile of a 
solitary wave solution (red curve) in a granular chain. One uses an arbitrary long-wavelength bell-shaped curve (blue curve) as initial data when solving Eq.~\eqref{strain}. \textbf{(b)} Contour plot of the evolution of the blue curve of panel (a) via
Eq.~\eqref{strain}. Observe that the high-amplitude portion of the wave (yellow) travels faster than the low-amplitude part (light blue). \textbf{(c)} Strain profile (at time $t=20$) showing the onset of microscopic oscillations, which are reminiscent of a dispersive shock. \textbf{(d)} Dispersive-shock solution of the KdV equation.
[Panel (d) is used with permission from \cite{scholar}.]
}
\label{fig:shock}
\end{figure}

\section{Dispersive Shock Waves} \label{sec:shocks}

As we discussed in Sec.~\ref{sec:qc} and Sec.~\ref{sec:more_qc}, one can derive a quasicontinuum approximation of a solitary wave solution in a long-wavelength limit of a monoatomic granular chain. However, a genuine solitary wave solution is strongly localized, with doubly-exponential tail decay when there is no precompression and exponential tail decay when there is precompression. If one considers a similar waveform, except with a considerably wider spatial scale (see Fig.~\ref{fig:shock}), it appears, upon evolving the initial data through Eq.~\eqref{strain}, that the wave speed of each point of the waveform depends on the amplitude at that point, with larger amplitudes featuring faster wave speeds (see Fig.~\ref{fig:shock}[b]). To understand this observation from an analytical perspective, let's consider the long-wavelength ansatz $y_n=Y(\eps n, \eps t) = Y(X,T)$. This yields
\begin{align} \label{ros2}
	\pa_T^2 Y = \pa_X^2[ (\delta_0 + Y)^p] \,. 
\end{align}
Equation \eqref{ros2} is reminiscent of Eq.~\eqref{pik}, except that now $\delta_0 \neq 0$ and there is no higher-order correction term.
We can rewrite equation \eqref{ros2} as a system of conservation laws \cite{dcdsa}:
\begin{equation} \label{psystem}
	\left .
		\begin{array}{cc}
			 \pa_T Y - \pa_X V &= 0\ \\
			\pa_T V - \pa_X [  (\delta_0 + Y)^p ] &=0\
		\end{array}
	\right\}\,,
\end{equation}
which have the form of a so-called ``$p$-system''~\cite{LeVeque,Smoller}.
Defining 
\begin{equation*}
	\mathbf{Y} = \begin{pmatrix}  Y \\ V  \end{pmatrix} \qquad \mathrm{and}  \qquad     F(\mathbf{Y} ) = \begin{pmatrix}  -V \\ -(\delta_0 + Y)^p   \end{pmatrix} 
\end{equation*}
we can express Eq.~\eqref{psystem} as a system of the Burgers type:
\begin{equation} \label{advection}
	 \pa_T \mathbf{Y} +  \mathrm{D}F(\mathbf{Y}) \pa_X \mathbf{Y} = 0\,,
\end{equation}
where $\mathrm{D}F$ 
\begin{equation*}
 	 \mathrm{D}F(\mathbf{Y}) = \begin{pmatrix}
 	   0  & -1 \\ -p(\delta_0+   Y)^{p-1} &0 
 	 \end{pmatrix}
\end{equation*}   
is the Jacobian matrix of $F$. The eigenvalues of $\mathrm{D}F(\mathbf{Y})$ are $ \lambda_{\pm}(\mathbf{Y}) = \pm \sqrt{ p( \delta_0 + Y)^{p-1}   } $, which suggests that solutions
of Eq.~\eqref{psystem} consist of traveling waves with velocity $\pm \sqrt{ p( \delta_0 + Y)^{p-1}   } $.
Because the velocity depends on the amplitude $Y$ of the solution (with larger speeds for
larger amplitudes if $p>1$), bell-shaped initial data deforms and steepens, as observed in Fig.~\ref{fig:shock}. For $\delta_0 \neq 0$, it was demonstrated rigorously in \cite{dcdsa} that Eq.~\eqref{ros2} accurately describes the dynamics
of a granular chain over long but finite time intervals. The approximation breaks down
when solutions of Eq.~\eqref{ros2} become singular (e.g., at the point of wave breaking).
In the inviscid Burgers equation $\pa_t u = \pa_x u^2$, the steepening of bell-shaped initial data leads ultimately to the formation of
``shock waves'' (i.e., front solutions that develop an infinite derivative in finite time) \cite{Drazin}.
Such derivative singularities cannot arise in a granular chain (because it is discrete in space), but the result of the amplitude-dependent wave speed (see Fig.~\ref{fig:shock}[b,c]) and subsequent steepening is reminiscent of systems that possess so-called ``dispersive shocks''~\cite{scholar}, in which microscopic oscillations spread out in space and time (see Fig.~\ref{fig:shock}[d]).  Something very similar occurs in granular chains with precompression \cite{dcdsa} (see Fig.~\ref{fig:shock}[c]), presumably
because of the dispersive role of discreteness when a structure becomes
sufficiently localized in space. Recently, dispersive shocks were studied numerically in FPU lattices with general convex potentials
for arbitrary Riemann (i.e., jump) initial data \cite{Herrmann10}.  
Because the $p$-system cannot describe the microscopic
oscillations observed in Fig.~\ref{fig:shock}(c), it is necessary to pursue other avenues to study them analytically. For example, in the presence of precompression, one can consider the KdV limit (see Eq.~\eqref{eq:kdv}) and derive Whitham 
equations to describe the amplitude of the oscillations of a KdV dispersive shock \cite{scholar}. 
A complementary perspective arises from identifying a first-order
equation (with $Y_T \propto (Y^{\frac{p+1}{2}})_X$) in a vein similar
to~\cite{mcdonald} and exploring its nonlinear transport properties.
One can also construct discretizations of the first-order equation and examine the properties for those discrete equations, analogous to the discretization analyses of Burgers equation in~\cite{wilma}. 
As our discussion suggests, the analytical study of dispersive shocks in granular
chains both with precompression and (especially) without precompression 
includes a considerable number of fascinating open questions.

Shock-like structures have been studied experimentally in homogeneous and periodic granular chains with $p=3/2$
and $\delta_0 = 0$ \cite{herbold,Molinari2009}. In those works, a ``shock wave'' was generated by applying a velocity to a single particle \cite{herbold} or by imparting velocity continuously to the end of a chain \cite{Molinari2009}. In these experimental works, the observed dynamics consists of a front that connects a large-amplitude state to a $0$-amplitude state in the displacements
(rather than the strain). They were thus termed ``shocks,'' but the dispersive shocks depicted
in Fig.~\ref{fig:shock} are in the strain variable and have not been explored experimentally.

\section{Breathers} \label{sec:breathers}


\subsection{Discrete Breathers and their Stability} \label{sec:breather_intro}

The final example of a prototypical nonlinear structure in a granular chain that we will discuss is a ``discrete breather'' (sometimes also called an ``intrinsic localized mode''), which is temporally periodic\footnote{Arguably, it may be desirable to loosen the definition and not demand strict periodicity.} and strongly localized in space \cite{PT_breather}. In the context of a granular chain, a discrete-breather (DB) solution of Eq.~\eqref{eq:model} satisfies the properties
\begin{equation*}
 	u_n(t) = u_n(t+T_b)\,,   \qquad   \lim_{n\rightarrow \pm \infty} | u_n - u_{n+1} |  = 0\,,
\end{equation*}	  
where $T_b$ is the breather period (see Fig.~\ref{fig:breathers}[a]). Breathers are generic excitations in spatially extended, discrete, periodic (or quasiperiodic) systems~\cite{Flach2007}. The span of systems in which such structures have been studied is broad and diverse: they include optical waveguide arrays and photorefractive crystals~\cite{moti}, micromechanical cantilever arrays~\cite{sievers}, Josephson-junction ladders~\cite{alex,alex2}, layered antiferromagnetic crystals~\cite{lars3,lars4}, halide-bridged transition metal complexes~\cite{swanson}, dynamical models of the DNA double 
strand \cite{Peybi}, Bose--Einstein condensates in optical lattices~\cite{Morsch}, and many others.

Once one finds a breather solution numerically (e.g., via a
fixed-point iteration of the time-$T_b$ map of a system
of interest (see Sec.~\ref{sec:numerics}), a natural followup question is its stability.
There have been extensive studies of the stability properties of breathers 
in discrete Hamiltonian systems (see \cite[4.2]{Flach2007} and  \cite{Aubry06,Flach05}). For
breathers in a granular chain, we focus on the notion of
``spectral stability,'' which is determined in the following way. One adds a perturbation to a breather solution and uses Eq.~\eqref{eq:model} to derive an equation that describes the evolution of the perturbation. Keeping only linear terms in this equation yields a Hill equation with temporally-periodic coefficients of period $T_b$
(see Eq.~\eqref{eq:var} of Sec.~\ref{sec:numerics}). Consequently, 
the eigenvalues (or Floquet multipliers) of the associated variational matrix $V(t)$ at time $t=T_b$ determine the spectral stability of breather solutions. Because of the Hamiltonian structure of the system, all Floquet multipliers (FMs) 
must lie on the unit circle for the solution to be (marginally) stable; otherwise, the solution is unstable (see Fig.~\ref{fig:breathers}[b]).
There are continuous arcs of spectrum on the unit circle (in the infinite-lattice limit), and one can compute these arcs from
the linear spectral bands of Eq.~\eqref{eq:model}. In general, the isolated multipliers (i.e., the ``point spectrum'') must be computed numerically (see Sec.~\ref{sec:numerics}). For Hamiltonian systems, there 
is always a pair of FMs at the point $+1$ (i.e., at the point $(1,0)$ of the unit circle). 
These correspond to the invariance of the system under time translation---a feature responsible for the conservation of the total energy.
 By exploiting information about this FM pair, one can extract a stability criterion for breathers~\cite{jcmprl} that is reminiscent of the well known Vakhitov--Kolokolov criterion for the stability of solitary waves. Specifically, this criterion establishes that a change of the monotonicity of the energy--frequency curve results in a change of stability of the corresponding breather solution.
 The above notion of stability is tantamount (for states such as traveling
 waves) to that of linear stability around stationary or traveling structures.
 The persistence of breathers under fully nonlinear evolution
 in granular chains (and in general nonlinear lattices) is still 
 incomplete. Therefore, at present, the above spectral stability predictions need to be validated through direct numerical simulation.

In the presence of damping, see Sec.~\eqref{sec:damped_driven}, the
picture is different, because then a granular chain is no longer Hamiltonian.
The continuous part of the spectrum now typically lies within the unit circle, so most eigendirections of perturbation decay over time. If all FMs lie within the
unit circle, the solution is not merely stable but it also
attracts nearby dynamics. Naturally, in this case, for a DB to exist
in the first place, there needs to be some form of forcing (or energy
pumping). Some of the examples that we will give in Sec.~\ref{sec:damped_driven} are in damped, driven granular chains. An interesting avenue for future work is to seek solutions in the form of dissipative discrete breathers (DDBs), which have been studied in systems such as RF superconducting quantum interference device (SQUID) arrays \cite{lazarides2007}.

\subsection{Nonexistence Results} \label{sec:nobreathers}

In the absence of precompression (i.e., when $\delta_{0,n} = 0$), the mean interaction force of a temporally-periodic solution $u$ must vanish: $\int_0^T V'( u_{n-1}(t) - u_n(t) ) dt = 0$. This observation, and the fact that $-V( u_{n-1} - u_n ) \leq 0$, implies that the only temporally-periodic solutions are the trivial equilibria (see \cite[Theorem 1]{James2013}). Consequently, for a homogeneous
chain, precompression is crucial for the emergence of DBs. The linear band structure that results from precompression allows a natural place from which breathers can bifurcate. In a monomer granular chain, one can try to construct approximate breather waveforms using a continuum modeling approach. Using the ansatz
\begin{equation} \label{nls_ansatz}
	y_n(t) \approx \eps Y( X,T) \exp(ik_0n + i\omega_0 t) + \mathrm{c.c.} + \mathrm{h.o.t.}\,, \qquad X = \eps( n - ct), \quad T = \eps^2 t\,,
\end{equation} 
where ``$\mathrm{c.c.}$'' is the complex conjugate and ``h.o.t.'' stands for higher-order terms, one can derive a nonlinear Schr\"odinger equation (NLS) from Eq.~\eqref{strain}.~\footnote{Note that the NLS ansatz \eqref{nls_ansatz} assumes small amplitudes, so one actually derives the NLS equation from the FPU model with a polynomial potential given by a Taylor expansion of the Hertz law (\ref{eq:taylor}). This is similar to the derivation of the KdV equation in Sec.~\ref{sec:preTW}.}. 
The NLS that one obtains is
\begin{equation} \label{nls}
	i \pa_T Y =   \nu_2 \pa_X^2 Y + B |Y|^2Y\,, \qquad \nu_2 = -\omega''(k_0)/2>0\,,  
\end{equation}
where $B$ is a lengthy wavenumber-dependent expression~\cite{Huang1,Schn10}. We also define the notation $\omega_0 = \omega(k_0)$, where $k_0$ is the wavenumber and $\omega(k_0)$ is the angular frequency given by the dispersion relation~\eqref{eq:disp} evaluated at the wavenumber $k_0$. We also introduce the notation $c = \omega'(k_0)$. The approximation~\eqref{nls_ansatz} represents a standing
DB if $c=0$ (i.e., when the dispersion curve has vanishing slope). For a granular chain,
this occurs at the edge of the spectrum (i.e., at $k_0=\pi$). At the wavenumber $k_0=\pi$, the coefficient $B$ is
\begin{equation} \label{eq:MI}
	B = 3 K_2 K_4 - 4 K_3^2\,.
\end{equation}
For any value of the nonlinearity exponent $p>1$, one obtains $B<0$ for a granular chain. 
To satisfy the spatial-localization criterion, the envelope function $Y$ must be spatially localized.
The soliton solution of the NLS equation~\eqref{nls} is a spatially localized function, but this solution exists only for the focusing NLS equation (i.e., when $B >0$). This suggests the possibility that breathers (of small amplitude and on top of a vanishing background) may not exist in a monomer granular chain. (As with other coherent structures and motivated by optical systems, breathers on top of a vanishing background are sometimes called ``bright'' breathers.) One can reach the same conclusion by realizing that linear plane wave solutions that evolve through the nonlinear equations of motion can collapse to a localized breather state through a modulational instability (MI) of a homogeneous background. For the monomer $K_2$--$K_3$--$K_4$ FPU lattice, the condition for MI of plane waves is $3 K_2 K_4 - 4 K_3^2>0$ \cite[Sec. 4.3]{Flach2007}. The expression on the left-hand side of this inequality is exactly the nonlinearity coefficient of the NLS equation (see Eq.~\eqref{eq:MI}). Because $B<0$, breathers cannot form through an MI. This discussion thus gives a heuristic argument that breathers do not exist. This result has been made mathematically rigorous using center manifold theory \cite{James03}, a rather different approach from the investigation of MI.
This result implies that to obtain a breather in a granular chain, one must modify the configuration either by altering the chain (e.g., with a heterogeneity or an on-site potential) or by seeking an alternate waveform (e.g., seeking a dark breather rather than a bright breather, as described in Sec.~\ref{sec:breather_variants}  ).  In the following subsections, we will explore both scenarios.

\subsection{Breathers in Chains with Impurities}  

By introducing an impurity (or ``defect'') into a precompressed (and hence linearizable) monomer granular chain, one can create a linear mode from which a breather can bifurcate. Such breathers are sometimes called ``impurity modes.'' It is possible to construct one or more breather branches as nonlinear continuations of (light-mass) defect modes of the linear problem. One can compute the eigenfrequencies and their modes as described in Sec.~\ref{sec:linear}.  If the defect mass is lighter than the other masses in the ``host'' chain, the corresponding eigenfrequency lies above the acoustic band, so the corresponding mode is spatially localized \cite{Craster2013}. 
 Otherwise, the eigenfrequency lies within the acoustic band, and the associated mode is spatially extended. In \cite{Theocharis2009}, it was shown using a granular chain with free boundary conditions and a single light defect that breathers can be continued in frequency from the localized linear defect mode to the edge of the acoustic band. The solutions feature a profile that is asymmetric. The case of two identical defects is qualitatively similar to the single-defect case if the defects are adjacent. However, if there is a particle between the two defects in a chain, there exists a linear defect mode with a symmetric profile and a linear defect  mode with an antisymmetric profile.  The branch corresponding to the continuation of the antisymmetric mode undergoes a symmetry-breaking pitchfork bifurcation in which two asymmetric solutions are created. If the two defects are different (e.g., in their masses), the antisymmetric branch undergoes an imperfect pitchfork bifurcation.

 \begin{figure} 
\centerline{
\includegraphics[width= \linewidth]{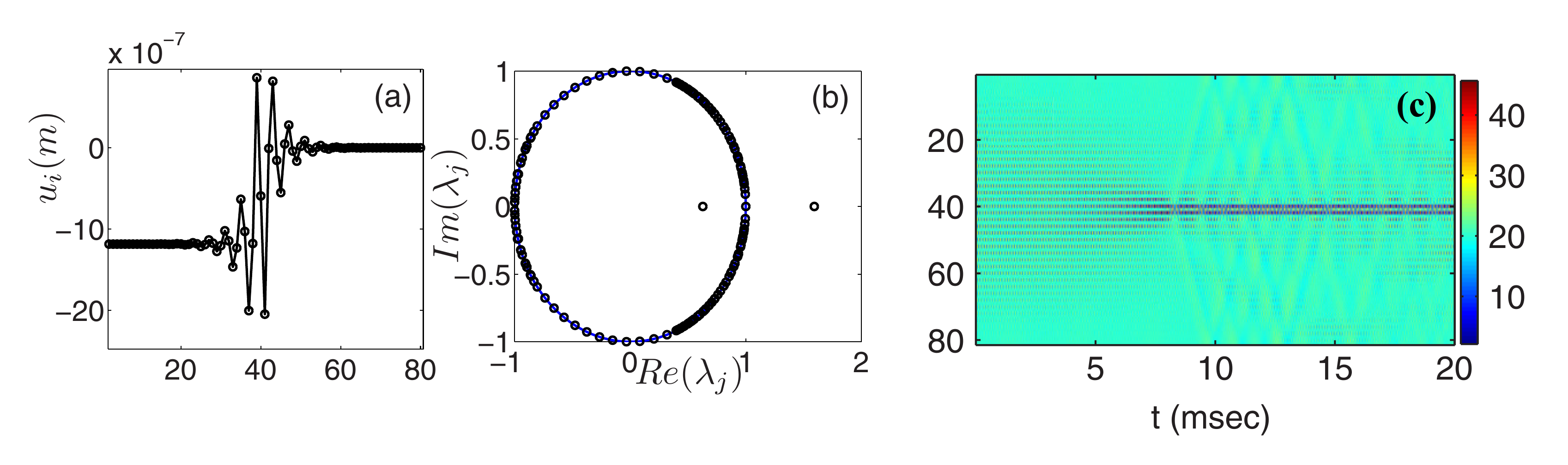}}
\caption{ (Color online). A Hamiltonian breather of a diatomic granular chain considered in \cite{Theocharis2009}.
\textbf{(a)} Profile in the displacement $u_n$ and \textbf{(b)} Floquet spectrum of the solution in panel (a). There is a single Floquet multiplier (FM) with modulus greater than $1$, indicating that there is an instability. \textbf{(c)} Spatiotemporal evolution of the
forces for a simulation of the modulation instability and ensuing breather generation from initial conditions that correspond to the lower optical cutoff (plane wave) mode (see Fig.~\ref{fig:dispersion}). [Panels (a,b) are adapted with permission from Ref.~\cite{Theocharis2009}. Copyrighted (2009) by the American Physical Society.
Panel (c) is adapted with permission from Ref.~\cite{Boechler2010}. Copyrighted (2010) by the American Physical Society.]
 }
\label{fig:breathers}
\end{figure}

\subsection{Breathers in Periodically Arranged Chains}

One obtains a periodic arrangement in an infinite granular chain by introducing an infinite number of defects in a periodic fashion. (The simplest example is a diatomic chain.) However, in contrast to a chain with a finite number of impurities, the linear mode
from which a breather bifurcates is not spatially localized. Instead, a DB forms spontaneously after a spatially extended linear plane wave undergoes an MI (see Fig.~\ref{fig:breathers}[c]).
In the context of granular chains, DBs were first studied theoretically and illustrated experimentally in~\cite{Boechler2010} using precompressed diatomic chains that consisted of alternating steel and aluminum spheres. The chain was driven at the boundaries at the first cutoff frequency of the optical band, and the corresponding mode experienced an MI.
This is because the bottom of the optical band of the dispersion satisfies $\omega''(k_0) >0$ (see Fig.~\ref{fig:dispersion}[a] )
 leading through its reduction to an effectively self-focusing NLS equation,\footnote{Recall, at the edge
of the spectrum $k_0=\pi$, the sole
branch of the dispersion relation in a monomer chain satisfies $\omega''(k_0) < 0$.}, which
is thus subject to the manifestation of modulation instability.
After some transient dynamics, a breather (spontaneously) forms, and it has a frequency that is located in the spectral gap that joins the acoustic and optical bands. One can obtain a more precise description of such a breather through numerical computations and subsequent parametric continuation. Several families of DB solutions in diatomic granular chains were studied in \cite{Theocharis10}. These breathers have different amplitudes and amounts of localization, and they can be either asymmetric or symmetric. The asymmetric breathers are centered at a light particle, and the symmetric ones are centered on a heavy particle. If the DB frequency is sufficiently close to the optical band, one can approximate the breather dynamics using the NLS equation \cite{Bruckner10,Huang2} (in a similar vein as we described in Sec.~\ref{sec:breather_intro}). The selection of both the frequency and the location of
the resulting breather when it emerges as a result of instability is still
an open problem. 

Breathers in chains with larger spatial periods have also been studied. For example, period-3 (i.e., ``trimer'') chains with sequences of 1 steel and 2 aluminum particles and period-4 (i.e., ``quadrimer'') chains with sequences of 1 steel and 3 aluminum particles were studied in~\cite{hooge11}. The trimer and quadrimer chains have, respectively,  two and three spectral gaps. A surprising, generic feature of the breathers in these systems is that the ones that bifurcate from the upper bands tend to be more robust (in terms of stability) than their counterparts that bifurcate from the lower bands. This appears to be because the former are able to avoid resonances with (higher-frequency) linear modes, but a systematic investigation of the stability of the various breathers remains an open problem.

 \subsection{Breather Variants in Damped, Driven Granular Chains} \label{sec:damped_driven}

To obtain a more realistic description of breathing solutions from experiments of granular chains, one needs to account for their external drive and damping. Perhaps the simplest damped, driven model involves replacing
one or both boundaries with a harmonic drive $u_0(t) = a \sin( 2 \pi f t) $, where $a,f \in \R$, and to add a (linear) dashpot term
$ - m_n/\tau \dot{u}_n$ to Eq.~\eqref{eq:model}, where the ``inverse damping strength'' $\tau$ (which is usually determined empirically) represents a time scale associated with the damping. 
One can interpret this form of dissipation as friction between individual grains and the supporting rods. 
This model for dissipation has been used for granular chains in several previous works \cite{Nature11,dark2,Stathis}, although (as noted in Sec.~\ref{subsec:model}) several studies (see, e.g., \cite{herbold,khatriprl,vergara,ortiz2012}) have considered internal friction caused by contact interaction between grains.

\begin{figure} 
\centering
\includegraphics[width=\linewidth]{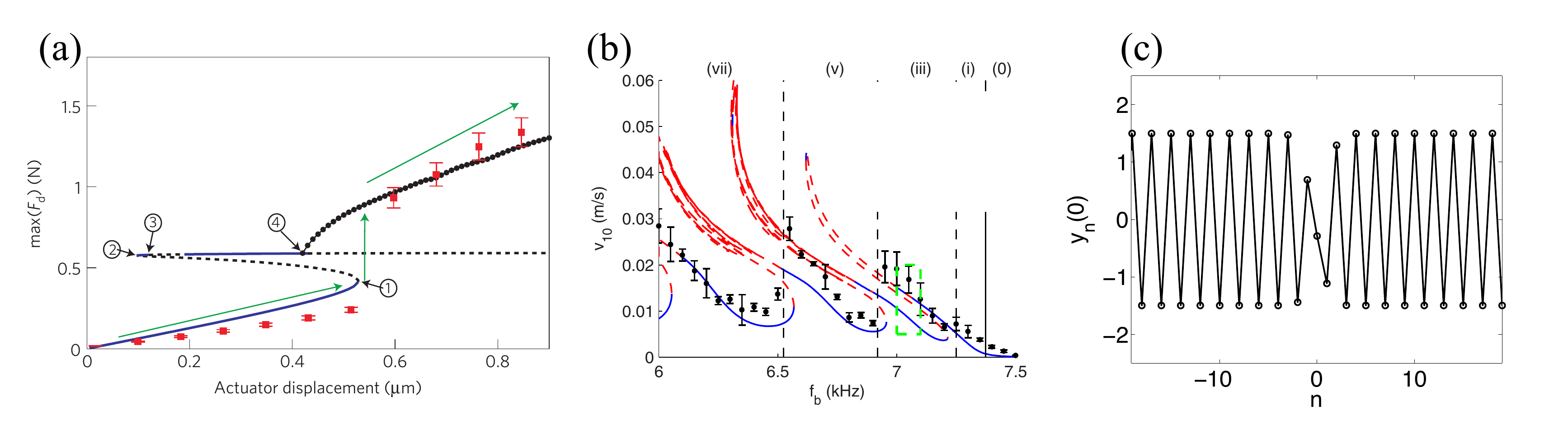} 
\caption{  \textbf{(a)} Continuation in driving amplitude of a damped, driven granular chain with an impurity near the drive source. We show the maximum force in the chain as a function of actuator displacement. Solid curves indicate stable branches, dashed curves indicate unstable branches, and dots with error bars correspond to experimentally-obtained points. \textbf{(b)} Continuation in driving frequency in a damped, driven monomer chain. We show the velocity of the 10th particle as a function of driving frequency. 
As in the previous panel, solid curves indicate stable branches, dashed curves indicate unstable branches, and dots with error bars correspond to experimentally-obtained points. 
\textbf{(c)} Dark breather in a monomer granular crystal. 
[Panel (a) is adapted with permission from Macmillan Publishers Ltd: \emph{Nature Materials}, Ref.~\cite{Nature11}, copyright (2010).
Panel (b) is adapted with permission from Ref.~\cite{dark2}. 
Copyrighted (2014) by the American Physical Society.
Panel (c) is reprinted with permission from Ref.~\cite{dark}.  
Copyrighted (2013) by the American Physical Society.]
} 
\label{fig:breathers2}
\end{figure}

Unlike in the associated Hamiltonian model, temporally-periodic solutions of a damped, driven granular chain can attract other types of dynamics, so one can obtain them by driving an initially at-rest chain for a sufficiently long time. These are the most straightforward breather-like solutions to
detect experimentally. For theoretical considerations and to obtain a more complete view
of the solution space of the system, one can also obtain unstable waveforms via Newton iterations (see Sec.~\ref{sec:numerics}).
One can study the resulting solutions as one varies the driving amplitude parameter $a$ or some other parameter.  Through numerical simulations and experiments, it has been shown in a chain with an impurity \cite{Nature11} and periodic arrangements (e.g., driven diatomic and triatomic chains~\cite{hooge12,Stathis}) that the interplay of nonlinear surface modes with modes caused by the driver create the possibility, as the driving amplitude is increased, of hysteretic dynamics and limit-cycle saddle--node bifurcations, beyond which the dynamics of the system becomes quasiperiodic or chaotic (see Fig.~\ref{fig:breathers2}[a]).  In light of these features, damped, driven granular chains have been proposed for acoustic switching and rectification applications \cite{Nature11}.  An intriguing phenomenon that is worth further investigation 
is that periodic orbits can arise even when the 
drive is applied to a chain without precompression (i.e., in the
sonic-vacuum regime). See \cite{vak1} for numerical study of this observation and \cite{vak2} for an experimental study.
Although the nonexistence result that we discussed in Sec.~\ref{sec:nobreathers} excludes the possibility of breathers in a purely nonlinear Hamiltonian granular chain, the works \cite{vak1,vak2} suggest that the presence of damping and driving allow the possibility of breathers in such chains. A rigorous proof 
of such a result is an open problem.

For a fixed, small driving amplitude (in particular, for $|a| \ll \delta_0$), one can perform a continuation in driving frequency to obtain a single branch that has peaks corresponding to the linear resonances. In general, as the driving amplitude increases, the peaks in the bifurcation diagram bend \cite{Lydon,dark2,Stathis} (see Fig.~\ref{fig:breathers2}[b]) in a way that is reminiscent of the 
foldover event of driven oscillators \cite{Mook}. See \cite{tristram} for a recent study that extends this idea to granular chains.  Such nonlinear bending can be useful for applications such as energy harvesting~\cite{eh}.

\subsection{Breather Variants in Homogeneous Chain Configurations} \label{sec:breather_variants}

As we highlighted in Sec.~\ref{sec:nobreathers}, the NLS equation~\eqref{nls}
in a homogeneous granular chain does not have bright-soliton solutions (on top of a vanishing background), because the nonlinear coefficient is negative (i.e., $B<0$).
In this case, the NLS equation is self-defocusing~\cite{siambook}, and it thus possesses dark-soliton solutions
\begin{align}
	 Y(X,T) = \sqrt{\kappa/B} \tanh\left( \sqrt{ \frac{-\kappa}{2 \nu_2 }}  X \right) e^{ i\kappa T}\,, \qquad \kappa <0\,, 
\end{align}
which have smaller amplitude than the background. Consequently, the approximation of
Eq.~\eqref{nls_ansatz} with the envelope function $Y$ chosen to be a dark soliton represents a breathing solution (with $k_0 = \pi$) on top of a nonzero background (i.e., it is a ``dark breather''): 
\begin{align}\label{approx_james}
	y_n(t) = 2 \eps (-1)^n   \sqrt{\frac{\kappa}{B}} \tanh \left( \sqrt{ \frac{-\kappa}{ 2 \nu_2 }}   \epsilon (n - x_0)   \right) \cos( \omega_b t )\,,
\end{align}
where $\omega_b = \omega_0 + \kappa \epsilon^2$ is the frequency of the breather (see Fig.~\ref{fig:breathers2}[c]). This approximation,
along with numerical computations, was used to study such states in a monomer granular chain in \cite{dark}
for frequencies that lie within the acoustic band. For frequencies close to the cutoff frequency $\omega_0$, the solutions are well approximated by the associated NLS description, and large-amplitude solutions bifurcate from the small-amplitude solutions described by the NLS equation. It was shown in \cite{dark} that spectral-stability predictions can fail because of the highly degenerate nature of monodromy matrices associated with these solutions for periodic boundary conditions. The experimental realization (and bistability in a damped, driven chain) of dark breathers
was studied in \cite{dark2}. An interesting byproduct of the investigations of bistability was the numerical and experimental identification of branches of solutions with 3, 5, 7, etc. dark breathers.

Localized breathing states can also arise in a granular chain without precompression if there is an additional on-site potential. Such a model can describe systems such as an array of cantilever beams decorated by spheres or, in
principle, a Newton's cradle under the influence of gravity \cite{James2013}. The breathers bifurcate from the lone linear mode introduced by the linear potential, and they tend to be more localized than the breathers that occur in granular chains with defects or periodic structures that we described in previous subsections. In this setting, the breathers (and traveling generalizations of them) are well described by the so-called ``discrete p-Schro\"odinger equation'' (DpS) \cite{James2011,James2014,James2013}, which one can derive through a multiple-scale analysis similar to ones that we have described throughout this review. This multiple-scale analysis has been justified rigorously (through suitable error bounds) in this setting \cite{James2011}.

Breathing states have also been explored in locally resonant homogeneous granular chains (see Eq.~\eqref{eq:localres} and Fig.~\ref{fig:local_res})~\cite{Liu2015,Liu2016}. In the presence of precompression, \cite{Liu2016} examined (i) traveling bright breathers and (ii) stationary and traveling dark breathers.
As with a monomer granular chain, one can use a multiple-scale expansion to derive an NLS equation---which is either of the defocusing or the focusing variety,
depending on the particular location in the acoustic or optical band---and thereby construct breathers.
Perhaps more surprisingly, in granular chains without precompression, periodic traveling waves and dark breathers can exist~\cite{Liu2015}, despite the absence of a linear spectrum from which these solutions can bifurcate. In this case, 
the relevant approximate model is a variant of the DpS equation (i.e.,
a model relying on averaging over the breather period, but which now has
a considerably more complicated form than the original DpS model) rather than an NLS equation.

\subsection{Numerical Computation of Breathers and their Spectral Stability}\label{sec:numerics}

We now provide a brief summary of how to compute breathers in 
granular crystals (as well as more generally) and how to assess their
stability using FMs. The primary tool used for the numerical construction of breathers is Newton 
fixed-point iterations \cite[Sec. 3]{Flach2007}.  
We begin by writing Eq.~(\ref{eq:model}) as a system of first order ODEs:
\begin{equation} \label{gc_compact}
	\dot{\mathbf{x}}=\mathbf{F}\left(t,\mathbf{u},\mathbf{v}\right)\,,
\end{equation}
with $\mathbf{x}=\left(\mathbf{u},\mathbf{v}\right)^{T}$, where 
$\mathbf{u}=\left(u_{1},\dots,u_{N}\right)^{T}$ and $\mathbf{v}=\dot{\mathbf{u}}$, respectively,
represent the $N$-dimensional position and velocity vectors. One then constructs a Poincar\'e map: $\mathbf{\cal{P}}(\mathbf{x}^{(0)})=\mathbf{x}^{(0)}-\mathbf{x}(T_{b})$,
where $\mathbf{x}^{(0)}$ is the initial condition and $\mathbf{x}(T_{b})$
is the result of integrating Eq.~(\ref{gc_compact}) forward in time until $t=T_{b}$ using standard ODE integrators \cite{Hairer}. A periodic solution with period $T_{b}$ (i.e., satisfying
the property $\mathbf{x}(0)=\mathbf{x}(T_{b})$) is then a root of the map 
$\mathbf{\cal{P}}$. To obtain this root, one applies Newton's method to the map 
$\mathbf{\cal{P}}$ in the following numerical-iteration step:
\begin{equation} \label{gc_newton_iter_scheme}
	\mathbf{x}^{(0,k+1)}=\mathbf{x}^{(0,k)}-\left[\mathcal{J}\right]^{-1}_{\mathbf{x}^{(0,k)}}\mathbf{\cal{P}}\left(\mathbf{x}^{(0,k)}\right)\,,
\end{equation}
where $k$ is the index of the Newton iteration and $x^{0}$ is the desired root of $\mathcal{P}$.  The Jacobian of the map $\mathcal{P}$ is $\mathcal{J}=\mathbf{I}-V(T_{b})$, where $\mathbf{I}$ is the $2N\times2N$ identity matrix; $V$ is the solution to the
variational problem
\begin{equation} \label{eq:var}
	\dot{V}=\left({\mathrm{D}F}\right)\,V\,,
\end{equation} 
with initial data $V(0)=\mathbf{I}$; and ${\mathrm{D}F}$ is the Jacobian of the 
equations of motion~(\ref{gc_compact}) evaluated at the point $\mathbf{x}^{(0,k)}$. 
One initiates the iteration step \eqref{gc_newton_iter_scheme} with a suitable guess (given, e.g., by a linear mode or a solution to a continuum nonlinear approximation) and applies it until one satisfies a user-prescribed 
tolerance criterion. One thereby obtains a temporally-periodic solution upon successful convergence (and with high accuracy). Because one must compute the monodromy matrix $V(T_{b})$ to
apply Newton iterations, one immediately obtains spectral stability information
by computing eigenvalues or FMs of $V(T_{b})$.  

For a Hamiltonian granular chain (i.e., Eq.~\eqref{eq:model} with no damping or driving), the monodromy
matrix $V(T_{b})$ has eigenvalues at $+1$ (see the discussion in Sec.~\ref{sec:breather_intro}), so the Jacobian
$\mathcal{J}=\mathbf{I}-V(T_{b})$ is not invertible. To overcome this issue, one must break the
degenerate nature of $\mathcal{J}$ by introducing additional constraints, such as a vanishing time average
(i.e., $ \int_0^T y_1(t) dt = 0$) or a pinning condition (e.g., $y_1(t) = 0$). In practice, it can also be effective to take a pseudoinverse of $\mathcal{J}$. See~\cite{Aubry06} and ~\cite{Flach2007} for numerous practical methods for computing the relevant periodic orbits.

\section{Two-Dimensional Granular Crystals} \label{sec:2D}

\subsection{Packing Geometries and Transient Dynamics}

\begin{figure}[t]
    \centering
    \includegraphics[width=\linewidth]{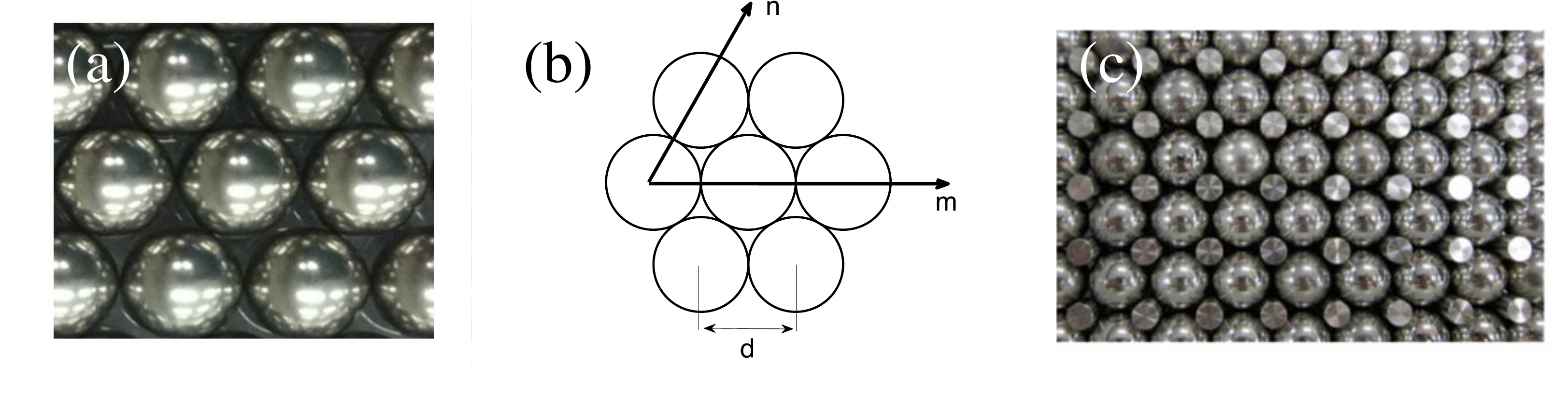}
 \caption{ \textbf{(a)} A granular crystal with hexagonal packing.  \textbf{(b)} Convention for the index orientation. The $m$-axis
 represents the horizontal direction, the $n$-axis represents the vertical direction, and the $m$-axis and $n$-axis meet at an
angle of $\theta = \pi/3$.  \textbf{(c)} A square granular lattice with cylindrical intruders. 
[Panels (a,c) were created by Andrea Leonard and are used with permission.
Panel (b) is reprinted with permission from Ref.~\cite{GCdirac}.]  
Copyrighted (2016) by the American Physical Society.
 }
 \label{fig:lattice}
\end{figure}

We now discuss 2D granular crystals, which have been investigated much less than the 1D configurations that we have discussed thus far. In two spatial dimensions, one can situate the nodes (i.e., particles) in a network in a large variety of ways, such as hexagonal (see Figs.~\ref{fig:lattice}[a,b]) or square lattices.  In a square lattice, one can generate a quasi-1D traveling solitary wave by exciting the lattice along one of the spatial dimensions. This is a trivial extension of Sec.~\ref{sec:TW}, as the waves never ``spill'' into adjacent directions. The situation for excitations in other directions is far less straightforward. However, from an experimental point of view, the square lattice is not robust structurally, because the particles can buckle readily, given the lack of contact with adjacent particles on the diagonals. One can overcome this issue by introducing ``intruder" particles into the vacant spaces between particles in a host crystal \cite{intruder1,ivan12} (see Fig.~\ref{fig:lattice}[c]). This is an experimentally relevant mechanism 
 for achieving either pulse redirection~\cite{ivan12} or (for intruders not restricted to local area in the crystal) 
 nonlinear pulse equipartition, a subject of
 intense theoretical investigation in granular crystals~\cite{intruder1}.
 In the experiment of~\cite{hasan2}, two rows of granular particles
 were interconnected via a chain of such interstitial particles.
 By exciting one of the rows (the ``impacted'' one), one can obtain an equal redistribution via the intruders between that row and
 the adjacent (or ``absorbing'') row.

Without additional sources to add energy to a granular crystal, one would not expect a genuine traveling solitary wave to exist in either a square or hexagonal lattice because the energy imparted to a particle upon impact is
distributed to an expanding array of neighbors as the pulse is transmitted. This process continues, leading gradually to attenuation
of the wave. Consequently, many papers on granular lattices of 2 or more dimensions are concerned with transient effects after a granular lattice (or more general packing) is struck \cite{Andrea,l6,l7,l10,l11,Aswasthi,Abd:2010,l9,Nishida,Coste:2008,Gilles:2003,Mouraille:2006,Leonard11}; 
this is relevant for impact mitigation applications. Additionally, structured granular composites, which are networks of 1D chains arranged in 3D space, exhibit exponential stress decay \cite{Leonard2015}.
From the perspective of applications, of particular interest is a ``sound bullet,'' in which a propagating wave front is directed to specific locations in a granular crystal~\cite{Spadoni}. One can also envision scenarios in which band gaps arise in 2D granular crystals either because of heterogeneities (e.g., impurities or periodic structures)
or local resonators (e.g., mass-in-mass systems).
In such scenarios, it would be very interesting to explore the existence, structure, 
and stability properties (and more generally, the temporal evolution) of localized states in the form of DBs or their variants (e.g., impurity modes, as we discussed for 1D configurations in Sec.~\ref{sec:breather_intro}).
In Fig.~\ref{fig:2D}(a), we show an example of a 2D localized mode of a 2D hexagonally packed granular lattice with a single light-mass defect in the center of the crystal. Another open problem is the study of phenomena such as ``quasipatterns'' and ``superpatterns,'' which have often been examined for settings such as Faraday waves in fluid systems and generic amplitude equations (e.g., complex Ginzburg--Landau equations with appropriate potentials) \cite{silber2002,silber2009}. 

\subsection{Hexagonal Granular Crystals: Equations of Motion, Linearized Dynamics, and Conical Diffraction}

\begin{figure}
     \centering
   \begin{tabular}{@{}p{0.33\linewidth}@{\quad}p{0.33\linewidth}@{\quad}p{0.33\linewidth}@{}}
      \subfigimg[width=\linewidth]{\bf (a)}{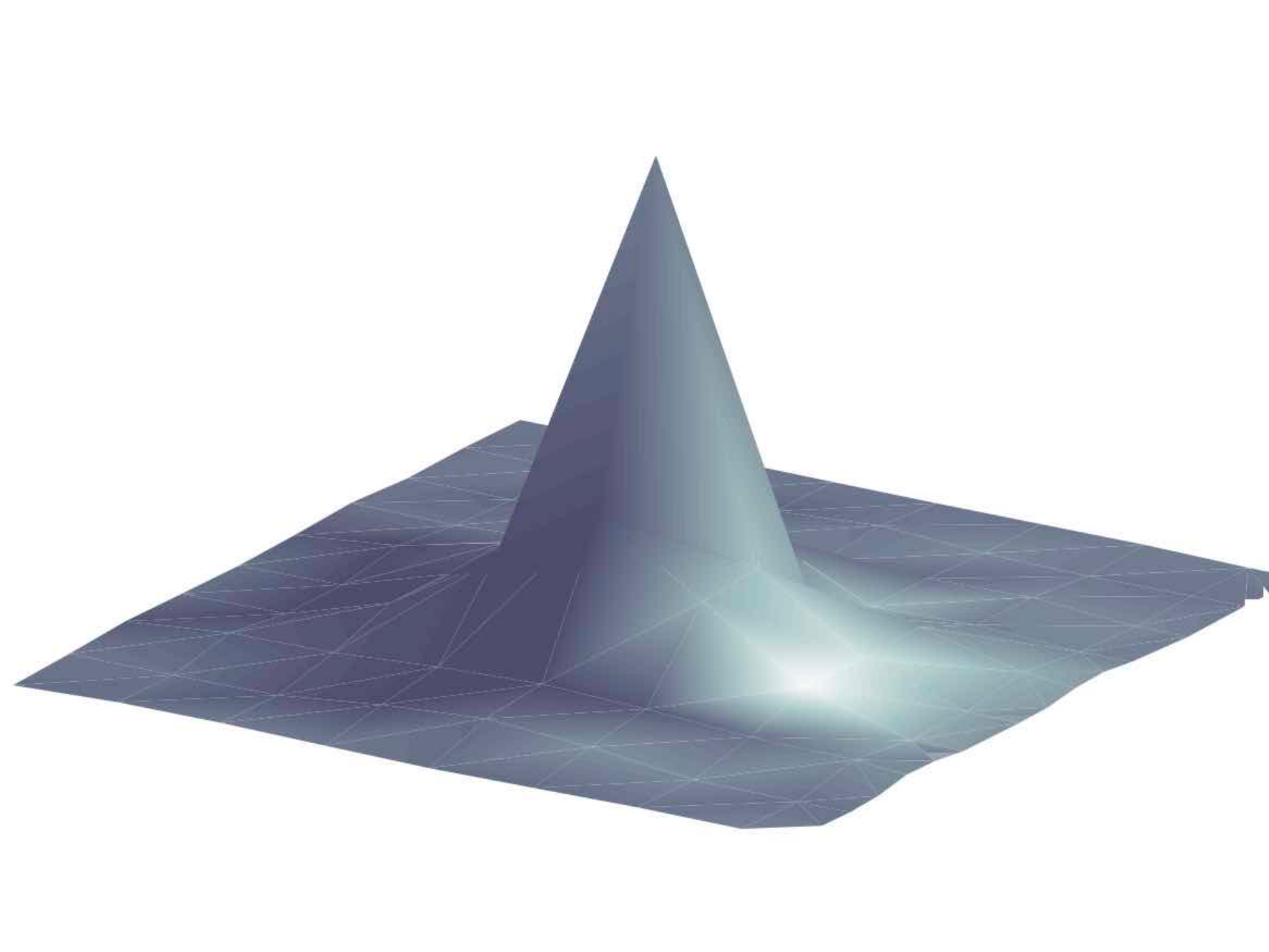}   &
      \subfigimg[width=\linewidth]{\bf (b)}{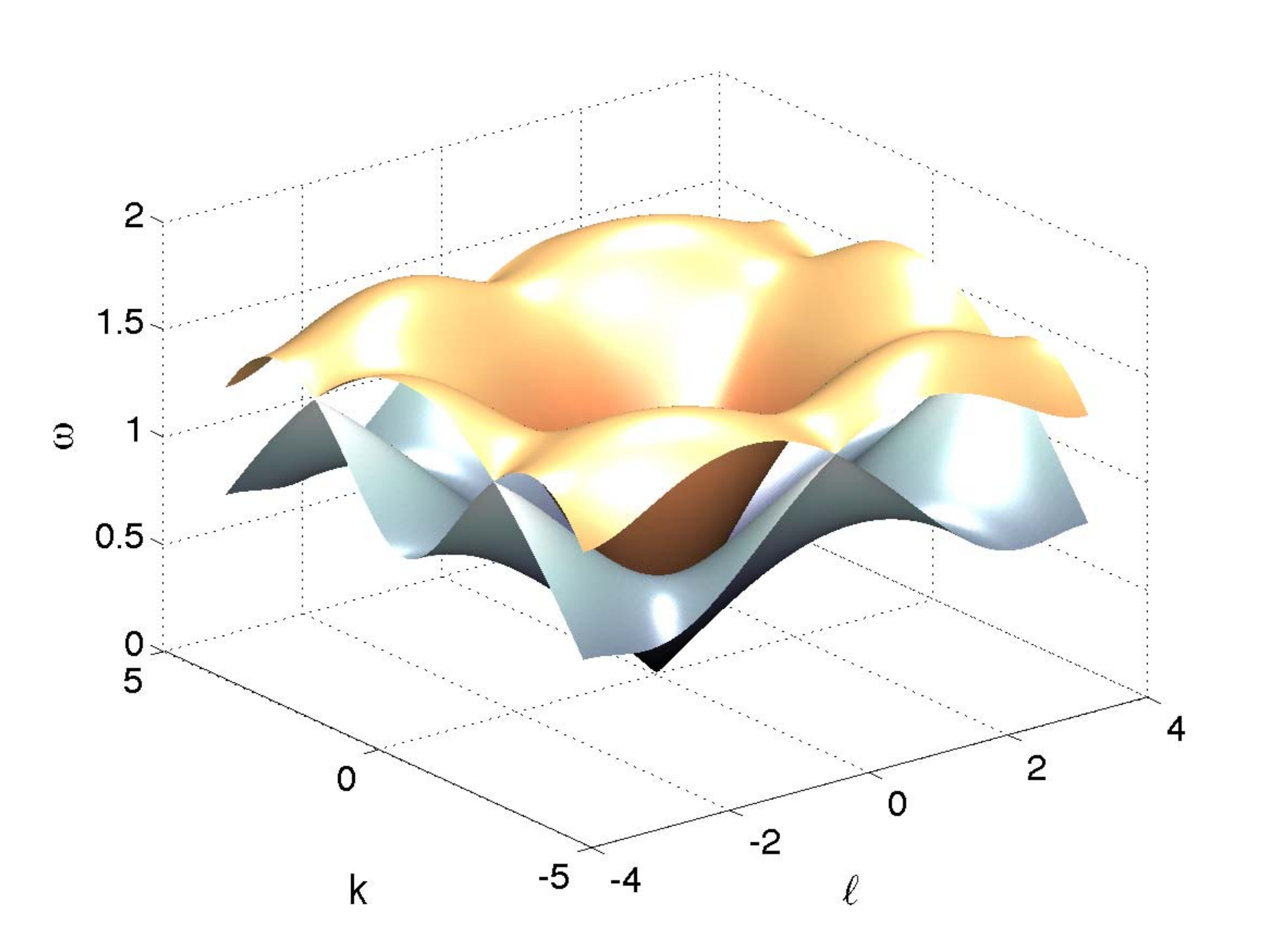}  &
      \subfigimg[width=\linewidth]{\bf (c)}{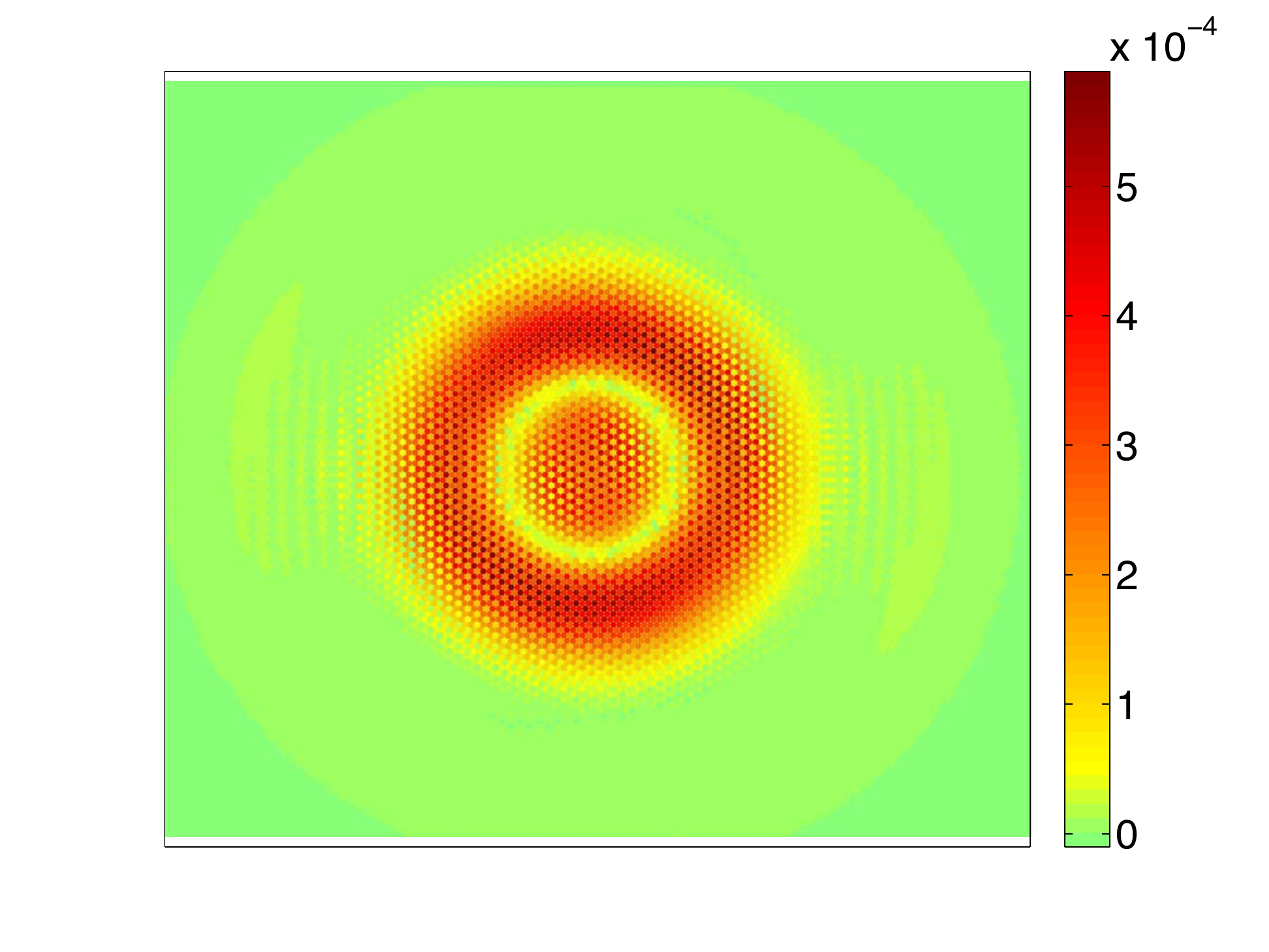} 
  \end{tabular}
 \caption{ \textbf{(a)} A breather solution of the 2D hexagonal granular network with a single defect. We show the magnitude of the displacement.
\textbf{(b)} Dispersion surface corresponding to Eq.~\eqref{eq:model2D}. The vector of wavenumbers is $(k,\ell)$, and $\omega$ is angular frequency. \textbf{(c)} Conical diffraction in the 2D hexagonal granular crystal described by equation \eqref{eq:model2D}. Such diffraction results from the excitation of linear modes near a Dirac point.
 [Panels (b,c) are reprinted with permission from \cite{GCdirac}. Copyrighted (2016) by the American Physical Society.]
}
\label{fig:2D}
\end{figure}

We now consider the specific case of a hexagonal granular crystal. The equations of motion account for the force at each contact point. They arise from Eq.~\eqref{eq:Hertz}, and each displacement $\mathbf{q}_{m,n}$ has both a horizontal component $u_{m,n}$ and vertical component $v_{m,n}$. If the granular crystal is precompressed, the equations are linearizable. In a hexagonal monomer lattice, the equations of motion are 
\begin{equation} \label{eq:model2D}
	\begin{array}{ll} \displaystyle
		\ddot{\mathbf{q}}_{m,n} =&  \mathbf{F}_1(\mathbf{q}_{m,n} -\mathbf{q}_{m-1,n})
		+ \mathbf{F}_2( \mathbf{q}_{m,n} - \mathbf{q}_{m,n-1})
		- \mathbf{F}_3(\mathbf{q}_{m+1,n-1} -\mathbf{q}_{m,n})
	\\
		& - \mathbf{F}_1(\mathbf{q}_{m+1,n} - \mathbf{q}_{m,n})  - \mathbf{F}_2(\mathbf{q}_{m,n+1} - \mathbf{q}_{m,n})
		+\mathbf{F}_3(\mathbf{q}_{m,n} - \mathbf{q}_{m-1,n+1})\,,
	\end{array}
\end{equation}
which takes into account the six contact points from the hexagonal symmetry. We let the mass $M = 1$ for notational simplicity. 
The vector-valued functions $\mathbf{F}_j(\mathbf{q}) = \mathbf{F}_j(u,v) = [ F_{j,u}(u,v), F_{j,v}(u,v) ]^T$ (with $j \in \{1,2,3\})$ have the form given in Eq.~\eqref{eq:Hertz} and take into account the relative positions of the particles in contact. For example, the force along the horizontal axis that results from contact between particles at positions $\{m,n\}$ and $\{m,n-1\}$ is
$F_{2,u}(u,v) =  A \left[ d - s \right]^{3/2}_{+} (d- \delta_0)( \cos(\theta) + u)/s$, where
$s =  \sqrt{ ( (d- \delta_0) \cos(\theta) + u )^2   + ( (d- \delta_0) \sin(\theta) + v  )^2}$,
the angle is $\theta = \pi/3$, the material parameter is $A$, and $d = 2R$ is the particle diameter. The nonlinearity strength is controlled by the term $\delta_0>0$, which represents how much the granular packing is precompressed (see Fig.~\ref{fig:lattice}[b]). Unlike many 2D systems (such as photonic systems \cite{Ablowitz1,Ablowitz2}), one can compute the dispersion 
relation explicitly (using a 2D discrete Fourier transform) instead of relying on numerical computations \cite{GCdirac}, see Fig.~\ref{fig:2D}(b).

In a hexagonal granular lattice, there exist so-called ``Dirac points'' in the dispersion surface at which upward-pointing and downward-pointing cones meet \cite{Hamilton,Lloyd}.  A notable example of another physical system that possesses Dirac points is graphene
 (a monolayer of graphite that exhibits considerable electron mobility~\cite{Novoselov1,Novoselov2}). One can thus consider a hexagonally packed granular crystal to be a ``phononic" analog of graphene; see also \cite{Torrent,Torrent2,Craster}. For wavenumbers near a Dirac point, one can derive so-called ``Dirac equations'' as a continuum model
using the methods described in this review \cite{GCdirac}.
The significance of this derivation is that ``conical diffraction'' (where the dynamics have the form of an expanding ring, as illustrated in Fig.~\ref{fig:2D}[c]) is possible in a Dirac system \cite{BorisBook}. Numerical
simulations of Eq.~\eqref{eq:model2D} validate the presence of conical diffraction in a hexagonal granular
crystal under small-amplitude excitations. Studies of conical diffraction (or refraction) date back nearly two centuries \cite{Hamilton,Lloyd}, and conical diffraction has also been studied
at length (both theoretically and experimentally) in photonic graphene \cite{Peleg,Ablowitz1}.
The presence of nonlinearity can play a crucial role in conical-diffraction dynamics,
and potentially it can even lead to a breakdown of conical
wave propagation in honeycomb lattices~\cite{fail,BorisBook}. The preliminary numerical
simulations of \cite{GCdirac} illustrate that conical diffraction can break down for significantly nonlinear responses. The study of the effects of nonlinearity on conical diffraction through, for example, the derivation of a nonlinear
Dirac system, is an open problem.  

\section{Conclusions and Outlook} \label{sec:theend}

Granular crystals are an exciting mechanical metamaterial, with the ability to shape the fundamental understanding of (both weakly and strongly) nonlinear systems and to test and inform numerous concepts from nonlinearity for applications. The study of granular crystals brings together challenging problems from applied mathematics, scientific computation, condensed-matter
and nonlinear physics (as well as other areas of physics), and experimental engineering. In this review, we have tried to illustrate the excitement that has been stimulated by recent advances in the study of some of the prototypical nonlinear structures in granular crystals: solitary waves, dispersive shocks, and breathers. 

A unifying theme of our review is approaches based on continuum approximations and asymptotics, which are illuminating in the study of all three families of structures. Although the continuum models provide important insights, they also suggest many open and nontrivial mathematical questions regarding their validity. (Even from a purely mathematical perspective, the equations themselves are intrinsically fascinating.) We have also argued that, in some cases, genuinely discrete methods may be better suited towards unraveling the dynamics of the wave structures. The numerical algorithms that are used for the identification of genuine traveling waves, and especially for stability computations, are at a very early stage.
The algorithm that we described for the numerical computation of breathers is adequate for 1D granular crystals (i.e., granular chains), but it is too inefficient for 2D crystals, creating the need for new and faster algorithms,
especially for large-scale computations of periodic orbits. The experimental
realization of phenomena such as conical diffraction and structures such as
the 2D breathers remain open, and a particular challenge arises from the need to incorporate important additional dynamical features, such as torsion and rotation, that play stronger roles in 2D systems than in 1D systems.
Throughout the review, we have also highlighted numerous other fascinating open issues. As we have discussed, granular crystals have a remarkable degree of tunability: one can consider particles with different masses, sizes, material properties, and geometries; one can consider different particle configurations (monomer chains, dimer chains, other periodic chains, chains with defects, and random configurations in 1D, and an even larger variety of possibilities in 2D); and one can tune precompression to examine crystals in either weakly nonlinear regimes or strongly nonlinear regimes, where the
latter is understood far more poorly than the former.
All of these features, along with numerous prototypes for applications---involving sensing, lensing, gating, impact mitigation, and vibration absorption, and more---render granular crystals an exciting test bed for the implementation
of linear and nonlinear, theoretical and applied, mathematical, physical, and engineering concepts. The study of granular crystals promises to be an exciting area for many years to come.


\section*{Acknowledgements}

The work of CC was funded partially by the NSF under Grant No. DMS-1615037.
PGK gratefully acknowledges the support of the US-AFOSR
under grant FA9550-12-1-0332 of the ERC under FP7; Marie
Curie Actions, People, International Research Staff Exchange
Scheme (IRSES-605096); the ARO under grant W911NF-15-1-0604;
and the Alexander von Humboldt Foundation.





\end{document}